\documentclass[12pt]{article}   	

\usepackage{geometry} 
\usepackage{amsmath} 								
\usepackage{graphicx}
\usepackage[font={small},labelfont={bf}]{caption}	
\usepackage{booktabs}
\usepackage{amssymb}
\usepackage{cite}
\usepackage{color,xspace}
\usepackage{xcolor}
\usepackage[hidelinks]{hyperref}
\usepackage{diagbox}
\usepackage{fancyvrb}
\usepackage{longtable}
\usepackage{multirow}
\usepackage{rotating}
\usepackage{slashed}
\usepackage{soul}
\usepackage{pdflscape}
\usepackage{authblk}
\usepackage{tabularx}
\usepackage{siunitx}

\numberwithin{equation}{section}
\DeclareSymbolFont{extraup}{U}{zavm}{m}{n}
\DeclareMathSymbol{\vardiamond}{\mathalpha}{extraup}{87}
\usepackage{float}
\restylefloat{table}
\allowdisplaybreaks

\usepackage{listings}
\definecolor{maroon}{cmyk}{0, 0.87, 0.68, 0.32}
\definecolor{halfgray}{gray}{0.55}
\definecolor{slha_frame}{RGB}{207, 207, 207}
\definecolor{slha_bg}{RGB}{247, 247, 247}
\definecolor{slha_red}{RGB}{186, 33, 33}
\definecolor{slha_green}{RGB}{0, 128, 0}
\definecolor{slha_cyan}{RGB}{64, 128, 128}
\definecolor{slha_purple}{RGB}{170, 34, 255}

\definecolor{mathematica_frame}{RGB}{207, 207, 207}
\definecolor{mathematica_bg}{RGB}{247, 247, 247}
\definecolor{mathematica_red}{RGB}{186, 33, 33}
\definecolor{mathematica_green}{RGB}{0, 128, 0}
\definecolor{mathematica_cyan}{RGB}{64, 128, 128}
\definecolor{mathematica_purple}{RGB}{170, 34, 255}

\lstnewenvironment{MIN}[1][]{%
  \renewcommand{\thelstnumber}{In[\arabic{lstnumber}]}
  \lstset{language=MathIn,numbers=left,basicstyle=\ttfamily,#1}%
}{%
}

\lstnewenvironment{MOUT}[1][]{%
  \renewcommand{\thelstnumber}{Out[\arabic{lstnumber}]}
  \lstset{language=MathOut,numbers=left,basicstyle=\ttfamily,#1}%
}{%
}

\usepackage{listings}
\lstdefinelanguage{SLHA}{
    morekeywords={block,Block,BLOCK,decay,Decay,DECAY},%
    sensitive=true,%
    morecomment=[l]\#,%
    morestring=[b]',%
    morestring=[b]",%
    morestring=[s]{'''}{'''},
    morestring=[s]{"""}{"""},
    morestring=[s]{r'}{'},
    morestring=[s]{r"}{"},%
    morestring=[s]{r'''}{'''},%
    morestring=[s]{r"""}{"""},%
    morestring=[s]{u'}{'},
    morestring=[s]{u"}{"},%
    morestring=[s]{u'''}{'''},%
    morestring=[s]{u"""}{"""},%
    identifierstyle=\color{black}\ttfamily,
    commentstyle=\color{slha_cyan}\ttfamily,
    stringstyle=\color{slha_red}\ttfamily,
    keepspaces=true,
    showspaces=false,
    showstringspaces=false,
    rulecolor=\color{slha_frame},
    frame=single,
    frameround={t}{t}{t}{t},
    framexleftmargin=6mm,
    numbers=left,
    numberstyle=\tiny\color{halfgray},
    backgroundcolor=\color{slha_bg},
    basicstyle=\footnotesize,
    keywordstyle=\color{slha_green}\ttfamily,
    aboveskip=1.2em,
    belowskip=1.2em,
}

\lstdefinelanguage{MathIn}{
    morekeywords={Simplify,Eigenvalues},%
    emph={Start,InitUnitarity,GetScatteringDiagrams,BuildScatteringMatrix,MakeSPheno},%
    emphstyle={\color{mathematica_purple}},
    sensitive=true,%
    morecomment=[l]\%,%
    morestring=[b]',%
    morestring=[b]",%
    morestring=[s]{'''}{'''},
    morestring=[s]{"""}{"""},
    morestring=[s]{r'}{'},
    morestring=[s]{r"}{"},%
    morestring=[s]{r'''}{'''},%
    morestring=[s]{r"""}{"""},%
    morestring=[s]{u'}{'},
    morestring=[s]{u"}{"},%
    morestring=[s]{u'''}{'''},%
    morestring=[s]{u"""}{"""},%
    identifierstyle=\color{black}\ttfamily,
    commentstyle=\color{mathematica_cyan}\ttfamily,
    stringstyle=\color{mathematica_red}\ttfamily,
    keepspaces=true,
    showspaces=false,
    showstringspaces=false,
    rulecolor=\color{mathematica_frame},
    frame=single,
    frameround={t}{t}{t}{t},
    framexleftmargin=10mm,
    numbers=left,
    numberstyle=\tiny\color{halfgray},
    backgroundcolor=\color{mathematica_bg},
    basicstyle=\footnotesize,
    keywordstyle=\color{mathematica_green}\ttfamily,
    aboveskip=1.2em,
    belowskip=1.2em,
}

\lstdefinelanguage{MathOut}{
    morekeywords={Simplify,Eigenvalues},%
    sensitive=true,%
    morecomment=[l]\%,%
    morestring=[b]',%
    morestring=[b]",%
    morestring=[s]{'''}{'''},
    morestring=[s]{"""}{"""},
    morestring=[s]{r'}{'},
    morestring=[s]{r"}{"},%
    morestring=[s]{r'''}{'''},%
    morestring=[s]{r"""}{"""},%
    morestring=[s]{u'}{'},
    morestring=[s]{u"}{"},%
    morestring=[s]{u'''}{'''},%
    morestring=[s]{u"""}{"""},%
    identifierstyle=\color{black}\ttfamily,
    commentstyle=\color{mathematica_cyan}\ttfamily,
    stringstyle=\color{mathematica_red}\ttfamily,
    keepspaces=true,
    showspaces=false,
    showstringspaces=false,
    rulecolor=\color{mathematica_frame},
    frame=single,
    frameround={t}{t}{t}{t},
    framexleftmargin=10mm,
    numbers=left,
    numberstyle=\tiny\color{halfgray},
    backgroundcolor=\color{mathematica_bg},
    basicstyle=\footnotesize,
    keywordstyle=\color{mathematica_green}\ttfamily,
    aboveskip=1.2em,
    belowskip=1.2em,
}

\let\origthelstnumber\thelstnumber
\makeatletter
\newcommand*\Suppressnumber{%
  \lst@AddToHook{OnNewLine}{%
    \let\thelstnumber\relax%
     \advance\c@lstnumber-\@ne\relax%
    }%
}

\newcommand*\Reactivatenumber{%
  \lst@AddToHook{OnNewLine}{%
   \let\thelstnumber\origthelstnumber%
   \advance\c@lstnumber\@ne\relax}%
}

\def\twomat[#1,#2][#3,#4]{\left( \begin{array}{cc} #1 & #2 \\ #3 & #4 \end{array} \right)}
\def\thv[#1,#2,#3]{\left( \begin{array}{c} #1 \\ #2 \\ #3 \end{array} \right)}
\def\twv[#1,#2]{\left( \begin{array}{c} #1 \\ #2 \end{array} \right)}

\newcommand{\met}{\ensuremath{E_\mathrm{T}^{\mathrm{miss}}}}

\newcommand{\ie}{\textit{i.e.}}

\newcommand{\madgraph}{{\sc MG5\_aMC}\xspace}

\newcommand{\madanalysis}{{\sc MadAnalysis\,5}\xspace}
\newcommand{\pythia}{{\sc Pythia\,8}\xspace}
\newcommand{\spey}{{\sc Spey}\xspace}
\newcommand{\pyhf}{{\sc PyHF}\xspace}
\newcommand{\hackanalysis}{{\sc HackAnalysis}\xspace}
\newcommand{\HackAnalysis}{{\sc HackAnalysis}\xspace}
\newcommand{\BSMArt}{{\sc BSMArt}\xspace}

\newcommand{\micromegas}{{\sc MicrOMEGAs}\xspace}
\newcommand{\checkmate}{{\sc CheckMATE}\xspace}

\newcommand{\orcid}[1]{\href{https://orcid.org/#1}{\includegraphics[width=10pt]{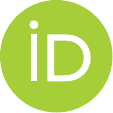}}}

\title{\textbf{A joint explanation for the soft lepton and monojet LHC excesses in the wino-bino model}}

\author[1]{Diyar Agin\thanks{\href{dagin.lpthe.jussieu.fr}{dagin@lpthe.jussieu.fr}}}
\author[1]{Benjamin Fuks\thanks{\href{mailto:fuks@lpthe.jussieu.fr}{fuks@lpthe.jussieu.fr}}}
\author[1]{Mark D. Goodsell\thanks{\href{mailto:goodsell@lpthe.jussieu.fr}{goodsell@lpthe.jussieu.fr}}}
\author[2]{Taylor Murphy\thanks{\href{murphyt6@miamioh.edu}{murphyt6@miamioh.edu}}}
\affil[1]{Laboratoire de Physique Th\'eorique et Hautes Energies (LPTHE),\protect\\ UMR 7589, Sorbonne Universit\'e \& CNRS\protect\\ 4 place Jussieu, 75252 Paris Cedex 05, France\vspace{2.5ex}}
\affil[2]{Department of Physics, Miami University\protect\\ 500 E. Spring St., Oxford, OH 45056, U.S.A.}

\date{}

\begin{document}

\maketitle

\abstract{
We present new recasts of the CMS Run 2 soft-leptons + missing energy analysis and the ATLAS Run 2 multijet + missing energy analysis. These analyses are relevant for probing the parameter space of electroweak-charged particles with compressed spectra. We review these analyses and detail their implementation and validation in \hackanalysis (for CMS) and \madanalysis (ATLAS). We then use these new recasts to combine four LHC analyses to identify a region of parameter space of the ``wino-bino'' simplified model, which corresponds to a limit of the Minimal Supersymmetric Standard Model in which higgsinos are decoupled, that is preferred over the Standard Model by excesses in the data. We find that the favoured region is compatible with the observed dark matter relic density, assuming freeze out within a standard cosmology, and we comment on the importance of this result and on how the simplified model should be mapped onto a complete supersymmetric model. 
}

\newpage
\setcounter{footnote}{0}

\tableofcontents



\section{Introduction}
\label{introduction}

Perhaps the most promising avenue for potential discoveries at the LHC in the coming years is to look for new particles with electroweak or electroweak-strength couplings. Such particles can still be produced copiously if they are not too heavy, but are often difficult to distinguish from Standard Model (SM) backgrounds; therefore there can be significant gains in sensitivity by adding new data and using new analysis techniques. For instance, the recent work in~\cite{Constantin:2025mex} includes a related discussion in the context of supersymmetry. 
Moreover, several experiments during Run 2 of the Large Hadron Collider (LHC) have already revealed tentative excesses \cite{ATLAS:2019lng,ATLAS:2021moa,CMS:2021edw,ATLAS:2021kxv,CMS:2021far} in searches for jets plus missing energy and soft leptons plus missing energy. In the series of papers \cite{Agin:2023yoq,Agin:2024yfs,Goodsell:2024aig,Fuks:2024qdt}, we showed that both signatures targeted by these searches could be produced by simple models, demonstrating that they overlap. In addition, we attempted to identify models and scenarios that would best fit the excesses, and we found preliminary evidence for minimal and next-to-minimal supersymmetric models with various specific light electroweakino compositions. 

Since supersymmetric models were the primary target of the soft-lepton searches, it is not surprising that the experimental collaborations have applied their results to intensive scans of parametrisations of such models known as the phenomenological Minimal Supersymmetric Standard Model (pMSSM). Notably, the ATLAS collaboration performed such a scan over the compressed electroweakino region in \cite{ATLAS:2024qmx}, where \emph{several} ATLAS analyses were applied to constrain the parameter space. The monojet or multijet + missing transverse energy ($\met$) searches were however ignored. Likewise, the CMS collaboration analysis \cite{CMS:2021edw} interpreted results in this way, and similar investigations have also been performed by theorists (\emph{e.g.} in~\cite{Altakach:2023tsd}). In these cases the focus has been on identifying the excluded regions rather than potential signals or models compatible with the excesses. 

Both ATLAS and CMS also use \emph{simplified models} to study the electroweak sector of low-energy supersymmetry. The idea behind simplified models is to ignore most supersymmetric particles, assuming that they are heavy, and focus on a representative subset of light particles assumed to be accessible at the LHC. This simplifies the presentation and interpretation of the results. For example, CMS used such simplified models to present combinations of electroweak searches \cite{CMS:2023qhl}. When discussing potential excesses this is a useful first step; in this work, we combine the analyses \cite{ATLAS:2019lng,ATLAS:2021moa,CMS:2021edw,CMS:2021far} for the first time, and therefore focus on such simple scenarios.

The simplified models relevant to the aforementioned excesses are known as ``higgsino'' and ``wino-bino'' scenarios, both corresponding to particular limits of the Minimal Supersymmetric Standard Model (MSSM). In the higgsino scenario, the starting point is a pair of pure higgsinos, \ie\ two $\mathrm{SU}(2)_{\text{L}}$ doublets with weak hypercharges $Y = \pm 1/2$, which together make two neutral Majorana fermions $\tilde{\chi}_{1,2}^0$ and one charged Dirac fermion $\tilde{\chi}_1^\pm$. It is then assumed that the masses of these fermions can be arbitrarily shifted away from each other (by mixing with the ignored heavier particles) without substantially affecting their interactions. The higgsino scenario then features $m_{\tilde{\chi}_1^\pm} = (|m_{\tilde{\chi}_2^0}| + |m_{\tilde{\chi}_1^0}|)/2$, with absolute values indicated because in the basis where the higgsino mixing matrices are real, the masses $m_{\tilde{\chi}^0_1}$ and $m_{\tilde{\chi}^0_2}$ have opposite sign. 
In the wino-bino scenario, the starting point is an $\mathrm{SU}(2)_{\text{L}}$ fermion triplet (wino) with zero weak hypercharge along with a gauge-singlet bino. Mixing between the bino and neutral wino is considered negligible, and the bino is assumed lighter than the winos. Again the spectrum includes two neutralinos and one chargino, but this time $m_{\tilde{\chi}_{1^\pm}} = |m_{\tilde{\chi}_2^0}|.$ There are two versions of this scenario, depending on whether the masses of the neutral fermions have the same or opposite signs. We shall focus on the case where $m_{\tilde{\chi}_2^0} \times m_{\tilde{\chi}_1^0} > 0$ (``wino-bino(+)'' in the parlance of the ATLAS collaboration) because the excess is more evident, and because it can include a dark matter (DM) candidate with properties compatible with cosmological data. In these simplified scenarios, the only possible light electroweakino interactions involve the $W$ and $Z$ bosons, and electroweakino decays in particular proceed via off-shell $W/Z$ boson exchanges according to
\begin{align}
    \tilde{\chi}_2^0 \rightarrow (Z^*\to f\bar f) + \tilde{\chi}_1^0\ \ \ \text{and}\ \ \ \tilde{\chi}_1^\pm \rightarrow (W^*\to f \bar f') +\tilde{\chi}_1^0,
\end{align}
with decay rates determined by kinematics. The simplified models therefore contain only two independent parameters: $m_{\tilde{\chi}_2^0}$ and $\Delta m \equiv |m_{\tilde{\chi}_2^0}| - |m_{\tilde{\chi}_1^0}|$.

Some of our previous works \cite{Agin:2023yoq,Agin:2024yfs,Goodsell:2024aig,Fuks:2024qdt} have examined the mild excesses in these CMS and ATLAS analyses in the context of higgsino and wino-bino simplified models. Interestingly, several of the favourable scenarios feature a lightest supersymmetric particle (LSP) that functions as a viable DM candidate, possessing a relic abundance (following freeze out during a standard cosmological history) within 20\% of the Planck collaboration value, $\Omega h^2 \approx 0.12$ \cite{Planck:2018vyg}, while remaining safe from direct-detection constraints. Until now, however, it has not been possible to conclusively identify the optimal region(s) of the simplified model parameter space owing to the lack of a reinterpretation of one of the experimental searches: the CMS Run 2 search for two soft leptons produced in association with missing transverse energy (CMS-SUS-18-004) \cite{CMS:2021edw}.

In this work, we close this gap by providing a reinterpretation code for this CMS analysis. Next, we apply it to the wino-bino(+) scenario, identifying the regions of the parameter space where the data describe the wino-bino model better than the SM and quantifying the excesses in all four analyses. We document this recast and its validation in Section~\ref{sec:RecastSoftLeptons}. We find that if the data are a harbinger of low-energy supersymmetry that can be described by the wino-bino model, then the wino most likely has a mass around $320$ GeV and the bino LSP has a mass around $20$~GeV beneath that. Moreover, we find that the excesses are best fit by a region of wino-bino parameter space that lies quite close to a band preferred by nearly pure bino DM with the correct relic abundance.

It has also been pointed out \cite{Buanes:2022wgm} that multijet-plus-missing energy searches can constrain the parameter space of these classes of models, even if they were designed to search for very different hypothetical particles. To test whether this will impact our results, we present in Section~\ref{sec:RecastMultijet} a recast of an ATLAS multijet-plus-missing energy search (ATLAS-SUSY-2018-22), and use it to probe compressed electroweakino models; for general interest, we also apply it to the compressed frustrated dark matter model presented in \cite{Fuks:2024qdt}. It is finally important to consider how representative simplified supersymmetric models actually are, \emph{i.e.}, under what circumstances genuine supersymmetric models will resemble simplified scenarios. We discuss these issues following the joint LHC statistical analysis in Section~\ref{sec:LimitInterpretation}, before concluding in Section~\ref{sec:Conclusion}.

\section{Recast of CMS-SUS-18-004 \texorpdfstring{(2/3$\boldsymbol{\ell + \met}$)}{2/3l+met} in \hackanalysis}
\label{sec:RecastSoftLeptons}

The CMS analysis CMS-SUS-18-004~\cite{CMS:2021edw} (henceforth ``CMS soft-lepton'') is broadly equivalent to the ATLAS analyses~\cite{ATLAS:2019lng,ATLAS:2021moa} in that they all target final states comprising ``soft'' leptons (having small transverse momentum) and missing energy along with at least one jet interpreted as originating from initial-state radiation. It supersedes (and greatly extends) the partial Run 2 analysis CMS-SUS-16-048 \cite{CMS:2018kag}. Until now there has been no publicly available recast code for the new CMS search, although a recast of the older search exists \cite{DVN/YA8E9V_2020,DVN/4WVPNQ_2020} in \madanalysis~\cite{Conte:2012fm,Conte:2014zja,Conte:2018vmg}. Unfortunately, the selections were sufficiently different as to make it impractical to reuse the older analysis code. In this section we present the details of the newer search and our recast implementation in \hackanalysis \cite{Goodsell:2021iwc,Goodsell:2024aig}.

The CMS soft-lepton search is split into two sets of regions intended to target different production mechanisms. One set, the ``$2\ell$ stop'' regions, targets production of colourful sparticles that then decay to electroweak charged ones; and the other ``EWK'' regions target pure electroweakino production. While the signal regions for the two cases are different, the control regions are the same and share a lot of features. However, we shall focus on the EWK regions here, as it is the one relevant for the excesses. 

\subsection{Selection regions}

We shall focus on the EWK selections for this analysis. 
They are divided into $2\ell$ and  $3\ell$ signal regions; two-lepton control regions named DY, $tt$ and SS; and a $3\ell$WZ region that is split into signal \emph{and} control regions. Each of these is are split according to the amount of adjusted \met: either just into Low ($\met \in [125, 200]$~GeV) or High ($\met > 200$~GeV) regions, or into Low ($\met \in [125, 200]$~GeV), Med ($\met \in [200, 240]$~GeV), High ($\met \in [240, 290]$~GeV) and Ultra ($\met > 290]$~GeV) regions. After the preselection cuts summarised in Table~\ref{TAB:Presel}, the regions are subdivided into bins corresponding to the value of the invariant mass of the dilepton pair $m_{\ell\ell}$. In the case of $3\ell$ regions the variable $M_{\rm SFOS}^{\min} (\ell\ell)$, the minimum invariant mass of all (at most two) same-flavour-opposite-sign lepton pairs, is used. This binning is summarised in Table~\ref{TAB:bins}. 
The bins are given a number in the associated recasting material/covariance matrix to label them ({\tt 0,1,2,3,4,5}), according to the $m_{\ell\ell}$ value; this numbering being somewhat counter-intuitive, we list it in the table. 

One of the curious features of this analysis is that the $3\ell$WZ region is split into signal and control regions. Here the bin with  $m_{\ell\ell} > 30$ GeV is indeed used as a control region, while the bins with lower $m_{\ell\ell}$ are used as signal regions. This stems from substantial signal contamination in the lower $m_{\ell\ell}$ bins: whereas they were initially intended to all be control regions only, it subsequentially made sense to split them and tag them as signal regions. In the table we list the bins defined by $m_{\ell\ell} \in [30,50]$ GeV with an asterisk: in the paper it is stated that there is no upper limit on $M_{\rm SFOS}^{\min} (\ell\ell)$ for the $3\ell$WZ regions, but in the recasting material the bins are listed as $m_{\ell\ell} \in [30,50]$ GeV. However, this distinction has little relevance for signals with compressed spectra. 


\begin{table}
\centering\renewcommand{\arraystretch}{1.25}
\resizebox{0.89\textwidth}{!}{\begin{tabular}{c|c c|c c c|c}\toprule\toprule
  Variable & SR $2\ell$   & SR 3$\ell$ & CR DY $2\ell$ & CR tt $2\ell$ & CR SS $ 2\ell$  &  3$\ell$WZ \\ \midrule
  $N_{\rm \ell} $ & 2 & 3 & 2 &2 & 2& 3 \\
  $p_{\text{T}} (\ell_i) $ for $e(\mu)$ [GeV] & \multicolumn{2}{c|}{$(5,30)$ (low), $(5(3.5),30)$ (high)} & \multicolumn{2}{c|}{$> 5$ (low), $> 5(3.5)$ (high)} & like SRs& --\\
  $p_{\text{T}} (\ell_1) > 30$ & \multicolumn{5}{c|}{--} & \checkmark \\
  $p_{\text{T}} (\ell_{2,3}) > 10 $  & \multicolumn{5}{c|}{--} & \checkmark \\
  $p_{\text{T}} (\ell_{2,3}) > 15 $& \multicolumn{5}{c|}{--} & $e$ (high only )\\
  $p_{\text{T}} > 20$ at least one $\mu$ & \multicolumn{5}{c|}{--} &  \checkmark \\
  1 OS pair & \checkmark & \checkmark & \checkmark & \checkmark   & --& \checkmark \\
  1 SS pair & -- & -- & -- & -- & \checkmark & -- \\
  1 OSSF pair & \checkmark & \checkmark & \checkmark & \checkmark  & --& \checkmark\\
  $\Delta R(\ell_i, \ell_j)$  &  \multicolumn{6}{c}{$>0.3$ (higher \met\ only)}  \\
  $M_{\rm SFOS}^{\rm min} (\ell \ell)$ & \multicolumn{5}{c|}{$\in [4,50]$~GeV [low], $\in [1,50]$~GeV [high]} & $> 4$~GeV, $>1$~GeV [l/h] \\
  $M_{\rm SFAS}^{\rm max} (\ell \ell)$  & -- & $< 60$ (low only) & -- & -- & -- & -- \\
  $M_{\rm SFOS}^{\rm min} (\ell \ell)$ veto $J/\Psi$, $\Upsilon$ &  \multicolumn{6}{c}{veto $(3,3.2)$ and $(9,10.5)$} \\
   $p_{\text{T}} (\ell\ell) $  & $>3$ & -- & $>3$ & $>3$ & $>3$  &  --  \\
   $j_1$ ``Tight lepton veto'' & \checkmark & -- & \checkmark & \checkmark & \checkmark&  --\\
  $m_{\text{T}}(\ell_i, p_{\text{T}}^{\rm miss} ), i=1,2 $ & $< 70$ &  -- & $< 70$ & -- & -- &-- \\
  $H_{\text{T}}$ &  \multicolumn{6}{c}{$>100$}\\
 $p_{\text{T}}^{\rm miss}/H_{\text{T}}$ & $(2/3,1.4)$ & -- & \multicolumn{3}{c|}{$(2/3,1.4)$} & -- \\
  $N_b(p_{\text{T}} > 25) $ & 0 & 0 & 0 & $\ge 1$ & 0 & 0  \\
  $M_{\tau\tau}$ & $<0$ or $> 160$ & -- & $(0,160)$ & \multicolumn{2}{c|}{$<0$ or $ > 160$} & -- \\
  Raw $p_{\text{T}}^{\rm miss} > 125$ & \multicolumn{6}{c}{all low regions} \\
  \bottomrule\bottomrule
  \end{tabular}}
  \caption{\label{TAB:Presel}Preselection criteria for CMS-SUS-18-004.}
\end{table}

In addition, we note that the preselection for the $2\ell$ stop region is identical to the $2\ell$ EWK \emph{except} that no OSSF requirement is enforced for the higher $m_{\ell\ell}$ regions (so that different lepton flavours are possible) while for the low $m_{\ell\ell}$ regions we still need muon pairs; and that no $m_{\text{T}}$ cut is implemented.

\begin{table}\centering
\captionsetup{width=0.85\textwidth}
\begin{tabular}{c c|c| c c c} \toprule\toprule
  Region & Variable & Low-MET & Med-MET & High-MET & Ultra-MET \\
         &          &   [GeV]      & [GeV] &         [GeV] &            [GeV] \\ \midrule
  SR $2\ell$ & $\met$ &$[125,200]$ & $[200,240]$ & $[240,290]$ & $> 290$ \\
  CR$2\ell$ & $\met$ & $[125,200]$ &  $> 200$  & -- & -- \\ \midrule
 All  $3\ell$ & $\met$ & $[125,200]$ &  $> 200$  & -- & -- \\ \midrule
  $m_{\ell \ell}$ 0 & $m_{\ell \ell}$& $[4,10]$ & \multicolumn{3}{c}{$[1,4]$} \\
  $m_{\ell \ell}$ 1 &  $m_{\ell \ell}$&$[10,20]$ & \multicolumn{3}{c}{$[4,10]$} \\
  $m_{\ell \ell}$ 2 &  $m_{\ell \ell}$&$[20,30]$ & \multicolumn{3}{c}{$[10,20]$} \\
  $m_{\ell \ell}$ 3 &  $m_{\ell \ell}$&$[30,50]^*$ & \multicolumn{3}{c}{$[20,30]$} \\
  $m_{\ell \ell}$ 4 &  $m_{\ell \ell}$&-- & \multicolumn{3}{c}{\,\,$[30,50]^*$} \\
  $m_{\ell \ell}$ 5 &  $m_{\ell \ell}$&-- & \multicolumn{3}{c}{--} \\
  \bottomrule\bottomrule
  \end{tabular}
  \caption{\label{TAB:bins} Binning for the signal and control-region bins of the EWK selections. The top three rows show the binning of the \met\ while the lower six rows show the subdivision into $m_{\ell\ell}$ bins.  The $m_{\ell \ell} \in [30,50]^*$ bin is listed with an asterisk because, as stated in the text, for the $3\ell$WZ regions there is an ambiguity whether they should be $[30,50]$ or $[30,\infty).$}
\end{table}



\subsection{Provided recasting material}

CMS provided several different pieces of information useful for recasting the analysis, in particular a covariance matrix for a simplified likelihood and detailed cutflows for specific benchmarks.
There are four cutflows for each of the two sets of regions defined in the analysis. For the electroweak signal regions, they correspond to the scenarios:
\begin{itemize}
\item {\tt TChi175/5}, representing a ``wino-bino'' model with $(m_{\tilde{\chi}_2^0}, \Delta m) = (175,5)$ GeV and a signal production cross section of 2.953 pb.
\item {\tt TChi200/40}, representing a ``wino-bino'' model with $(m_{\tilde{\chi}_2^0}, \Delta m) = (200,40)$ GeV and a signal production cross section of 1.807 pb.
\item {\tt Higgsino160/3}, representing a ``higgsino'' model with $(m_{\tilde{\chi}_2^0}, \Delta m) = (160,3)$ GeV and a signal production cross section of 1.539 pb.
\item {\tt Higgsino180/10}, representing a ``higgsino'' model with $(m_{\tilde{\chi}_2^0}, \Delta m) = (160,3)$ GeV and a signal production cross section of 1.001 pb.
\end{itemize}
The {\tt TChiWZ} models are \emph{pure wino} scenarios with degenerate $\tilde{\chi}_2^0$ and $\tilde{\chi}_1^\pm$ decaying via offshell $W/Z$ bosons only, and for which solely $\tilde{\chi}_2^0\tilde{\chi}_1^\pm$ pair production is considered. For the {\tt Higgsino} models, the condition $m_{\tilde{\chi}_1^\pm} = (m_{\tilde{\chi}_1^0} + m_{\tilde{\chi}_2^0})/2$ is imposed as for the corresponding ATLAS interpretations, and both $\tilde{\chi}_2^0\tilde{\chi}_1^\pm$ and $\tilde{\chi}_2^0 \tilde{\chi}_1^0$ production is considered for signal production, \emph{but not chargino pair production}. These correspond to the two main scenarios considered in the analysis paper, although CMS also considers a pMSSM higgsino model that is not relevant for the chosen benchmarks.

There was no efficiency information provided for this analysis, but a set of efficiency information for the reconstruction of electrons and muons was provided for older analyses at \url{https://twiki.cern.ch/twiki/bin/view/CMSPublic/SUSMoriond2017ObjectsEfficiency}; of relevance to us in the older (much smaller luminosity) version of this analysis, CMS-SUS-16-048 \cite{CMS:2018kag}. We must assume that these efficiencies are still relevant, and indeed they appear to correspond well to remarks made in the analysis paper. 

Two covariance matrices for simplified likelihoods were provided for this analysis, one for the EWK and one for the stop regions. The signal regions are divided into three years, corresponding to the data-taking periods. However, the individual yields for background and observed events are only provided in aggregates across all years; and in any case it is difficult to model the subtle differences between the years since the main difference is in the trigger requirements. Hence we combine the years together in our recast implementation. This is straightforward for simplified likelihoods following the procedure described in \cite{Collaboration:2242860}\footnote{\url{http://cds.cern.ch/record/2242860/}} Section 3 (``aggregated signal regions''). The result is that when combining bins $\{i\}$ into a smaller set of bins $\{I\}$, where a bin from $I$ is formed by combining certain bins of $\{i\}$ organised in a set $c_I$, then the means add. More specifically, if the expected number of background events in bin $i$ is $n_i$ and in the combined bins is $n_I$, then
\begin{align}\label{eq:agreeg1}
n_I = \sum_{i \in c_I} n_i.
\end{align}
For instance, if we combine bins $1$ and $2$ into a new bin $A$, then $c_A = \{ 1,2\}$ and the mean number of events in the combined bin is $n_A = n_1 + n_2$. Similarly, we combine the elements of the covariance matrix $V_{ij}$ into new elements $V_{IJ}$ as
\begin{align}\label{eq:agreeg2}
V_{IJ} = \sum_{i \in c_I} \sum_{j \in c_J} V_{ij}.
\end{align}
Note that this pertains to the covariance and not the correlation matrix. 
For any bins that are not combined then $c_I$ contains only one element and the above formulae hold.

Using the above, we produced a covariance matrix and {\tt info} file with the necessary statistical information that we subsequently used in \HackAnalysis and could also, in future, be used in {\sc MadAnalysis\,5}. 

\subsection{Validation}
\label{sec:CMSValidation}

The analysis was implemented in \hackanalysis and is now included in version 2.4\footnote{\url{https://goodsell.pages.in2p3.fr/hackanalysis/}}.

To validate our implementation, we first compared our predictions for the four electroweak cutflows provided by CMS and listed above.
Unfortunately, the experimental cutflows were clearly generated with bias at the generator level, with no indication as to what the bias was. Therefore it is impossible to follow the flow of cuts. On the other hand, since cross sections for each point were provided\footnote{Corresponding to tabulated values found at \url{https://twiki.cern.ch/twiki/bin/view/LHCPhysics/SUSYCrossSections}}, it is possible to infer the final efficiencies for each signal region and compare. Since, in our simulations, we force the decay $\tilde{\chi}_2^0 \rightarrow \tilde{\chi}_1^0 \ell^+ \ell^-$, we compute the total number of predicted events including a factor of the branching ratio. We then compare the values in Table~\ref{TAB:Higgsino} for the two higgsino points and Table~\ref{TAB:WinoBino} for the two wino-bino points. Here, the simulations are performed without bias or generator-level cuts (except for forcing the heavy neutralino decay in order to capture the correct $m_{\ell\ell}$ distribution; see \cite{Agin:2024yfs} for a discussion), and rely on leading-order matrix elements generated by \textsc{MadGraph5\texttt{\textunderscore}aMC@NLO} version 3.5.1 (\madgraph)~\cite{Alwall:2014hca}, using the UFO~\cite{Degrande:2011ua, Darme:2023jdn} model generated in \cite{Duhr:2011se} with \textsc{FeynRules}~\cite{Christensen:2009jx, Alloul:2013bka}. Moreover, these matrix elements are convoluted with the leading order set of {\sc NNPDF} v3.1 \cite{NNPDF:2017mvq} parton distribution functions accessed via {\sc LHAPDF} \cite{Buckley:2014ana}, and feature up to two additional jets in the final state that we combine following the MLM scheme~\cite{Mangano:2006rw, Alwall:2008qv}, before showering them with \pythia~\cite{Bierlich:2022pfr}. This then yields over a million events after merging. Agreement can be seen to be very good. However, while the signal regions are subsequently broken down into $m_{\ell\ell}$ bins, the individual yields/efficiencies were not provided for these, nor were yields provided for the control regions or the WZ signal regions despite these being included in the simplified likelihood provided by CMS.

\begin{table}\renewcommand{\arraystretch}{1.2}\centering
\resizebox{0.485\textwidth}{!}{\begin{tabular}{c|c c} \toprule\toprule
Signal region & Experiment & HackAnalysis \\ \midrule
$2\ell$ SR High-MET & $ 2.9\times 10^{-5} $ & $ 1.9^{+0.36}_{-0.36}\times 10^{-5} $\\
$2\ell$ SR Low-MET & $ 1.4\times 10^{-6} $ & $<$ $ 3.2\times 10^{-6} $\\
$2\ell$ SR Med-MET & $ 1.7\times 10^{-5} $ & $ 1.6^{+0.31}_{-0.31}\times 10^{-5} $\\
$2\ell$ SR Ultra-MET & $ 7.5\times 10^{-5} $ & $ 5.7^{+0.63}_{-0.63}\times 10^{-5} $\\
$3\ell$ SR Low-MET & $ 0.0 $ & $<$ $ 3.2\times 10^{-6} $\\
$3\ell$ SR Med-MET & $ 0.0 $ & $ 1.2^{+0.87}_{-0.87}\times 10^{-6} $\\
\bottomrule\bottomrule
\end{tabular}}\hfill 
\resizebox{0.485\textwidth}{!}{\begin{tabular}{c|c c} \toprule\toprule
Signal region & Experiment & HackAnalysis \\ \midrule
$2\ell$ SR High-MET & $ 4.2\times 10^{-4} $ & $ 3.2^{+0.13}_{-0.13}\times 10^{-4} $\\
$2\ell$ SR Low-MET & $ 3.7\times 10^{-4} $ & $ 4.1^{+0.14}_{-0.14}\times 10^{-4} $\\
$2\ell$ SR Med-MET & $ 5.6\times 10^{-4} $ & $ 3.6^{+0.13}_{-0.13}\times 10^{-4} $\\
$2\ell$ SR Ultra-MET & $ 6.3\times 10^{-4} $ & $ 6.4^{+0.18}_{-0.18}\times 10^{-4} $\\
$3\ell$ SR Low-MET & $ 1.3\times 10^{-5} $ & $ 4.0^{+0.44}_{-0.44}\times 10^{-5} $\\
$3\ell$ SR Med-MET & $ 8.4\times 10^{-5} $ & $ 9.9^{+0.70}_{-0.70}\times 10^{-5} $\\
\bottomrule\bottomrule
\end{tabular}}
\caption{\label{TAB:Higgsino}Comparison of CMS {\tt Higgsino160/3} (left) and {\tt Higgsino180/10} (right) efficiencies.}
\end{table}

\begin{table}\renewcommand{\arraystretch}{1.2}\centering
\resizebox{0.485\textwidth}{!}{\begin{tabular}{c|c c} \toprule\toprule
Signal region & Experiment & HackAnalysis \\ \midrule
$2\ell$ SR High-MET & $ 1.1\times 10^{-4} $ & $ 1.7^{+0.09}_{-0.09}\times 10^{-4} $\\
$2\ell$ SR Low-MET & $ 5.4\times 10^{-4} $ & $ 6.2^{+0.17}_{-0.17}\times 10^{-4} $\\
$2\ell$ SR Med-MET & $ 2.6\times 10^{-4} $ & $ 1.9^{+0.09}_{-0.09}\times 10^{-4} $\\
$2\ell$ SR Ultra-MET & $ 1.3\times 10^{-4} $ & $ 1.7^{+0.09}_{-0.09}\times 10^{-4} $\\
$3\ell$ SR Low-MET & $ 2.8\times 10^{-4} $ & $ 2.7^{+0.11}_{-0.11}\times 10^{-4} $\\
$3\ell$ SR Med-MET & $ 3.6\times 10^{-4} $ & $ 2.9^{+0.11}_{-0.11}\times 10^{-4} $\\
\bottomrule\bottomrule
\end{tabular}}\hfill
\resizebox{0.485\textwidth}{!}{\begin{tabular}{c|c c}
\toprule\toprule
Signal region & Experiment & HackAnalysis \\ \midrule
$2\ell$ SR High-MET & $ 8.3\times 10^{-5} $ & $ 1.1^{+0.08}_{-0.08}\times 10^{-4} $\\
$2\ell$ SR Low-MET & $ 3.7\times 10^{-5} $ & $ 3.9^{+0.47}_{-0.47}\times 10^{-5} $\\
$2\ell$ SR Med-MET & $ 1.5\times 10^{-4} $ & $ 1.0^{+0.07}_{-0.07}\times 10^{-4} $\\
$2\ell$ SR Ultra-MET & $ 3.1\times 10^{-4} $ & $ 2.8^{+0.13}_{-0.13}\times 10^{-4} $\\
$3\ell$ SR Low-MET & $ 2.9\times 10^{-6} $ & $ 7.5^{+2.09}_{-2.09}\times 10^{-6} $\\
$3\ell$ SR Med-MET & $ 6.3\times 10^{-5} $ & $ 5.0^{+0.56}_{-0.56}\times 10^{-5} $\\
\bottomrule\bottomrule
\end{tabular}}
\caption{\label{TAB:WinoBino}Comparison of CMS {\tt TChi200/40} (left) and {\tt TChi175/5} (right) efficiencies.}
\end{table}

\begin{figure}
\captionsetup{width=0.85\textwidth}
\centering
  \includegraphics[width=0.6\textwidth]{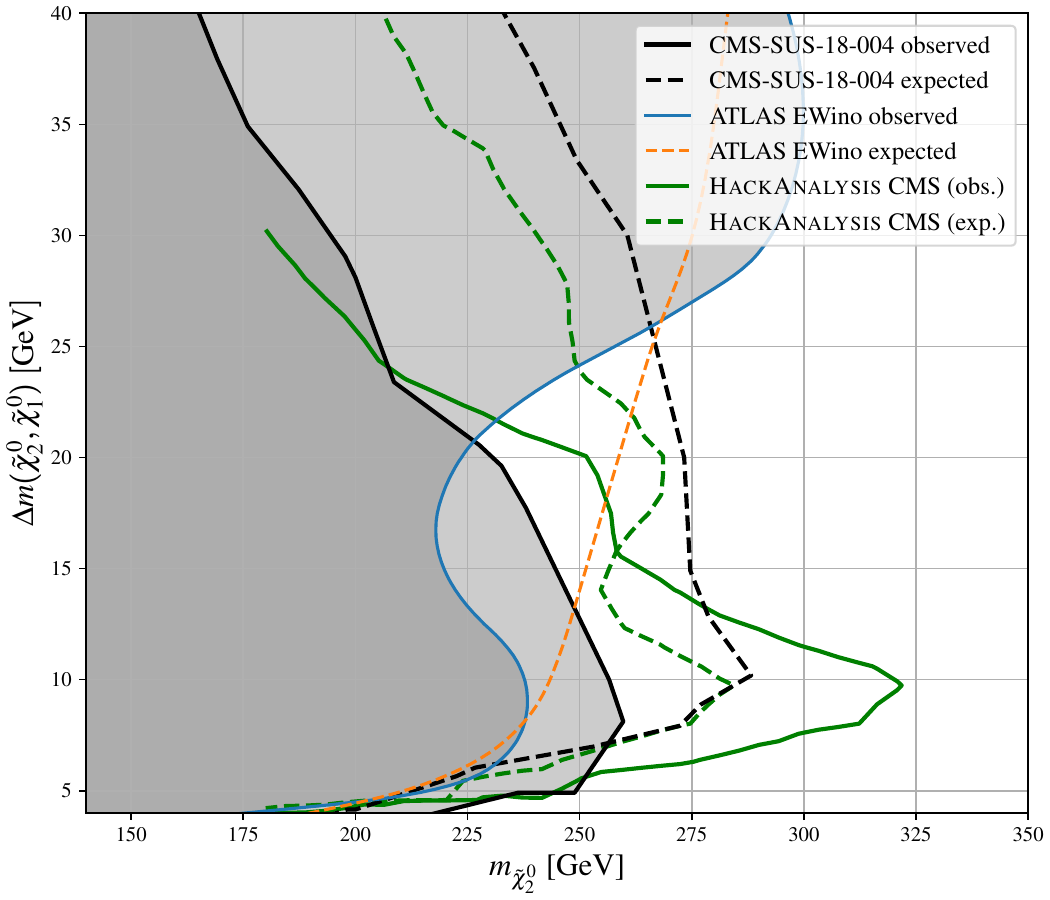}\\[1cm]
  \includegraphics[width=0.6\textwidth]{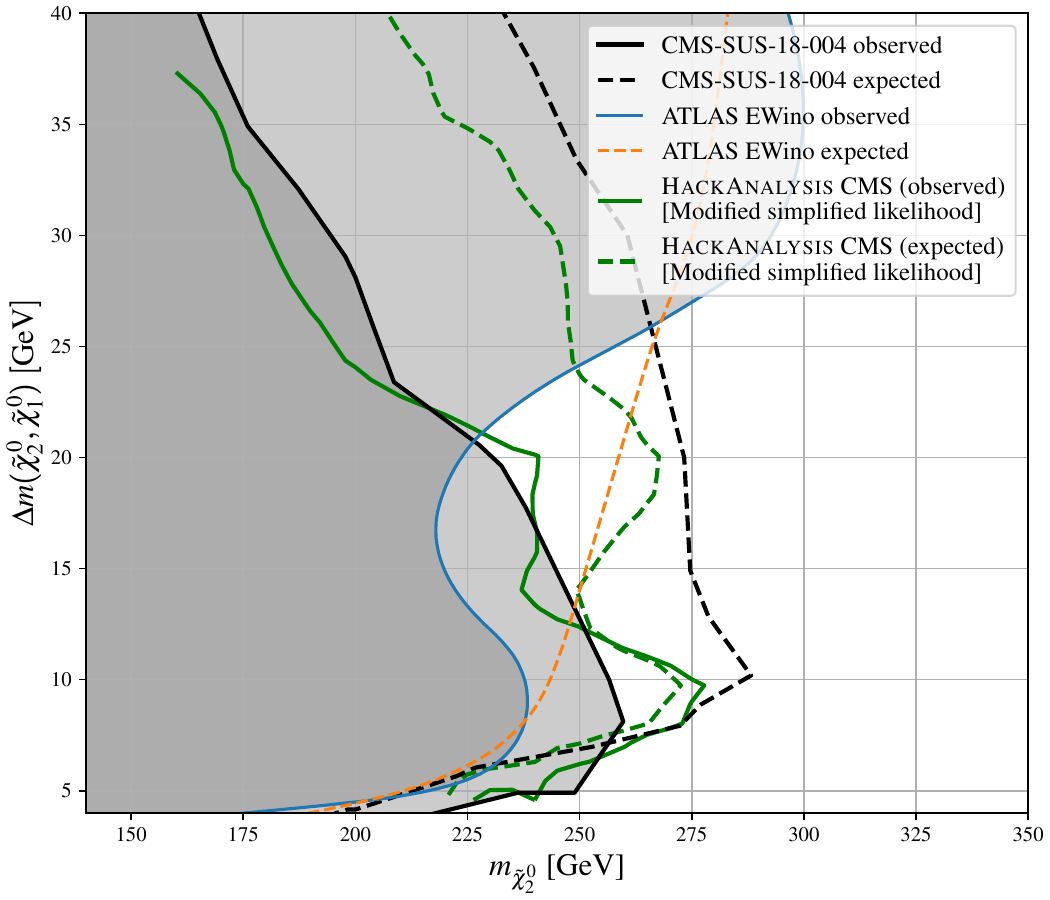}
  \caption{\small\label{FIG:CMS2Llimits}Validation of the implementation of CMS-SUS-18-004 in \hackanalysis by comparing limits on the wino-bino parameter space. Top: ``Full'' simplified likelihood. Bottom: Alternative simplified likelihood, by combining two pairs of bins. See text for details.}
\end{figure}

We then turn to a validation based on the reproduction of exclusion curves, focusing on the wino-bino scenario with same-sign mass eigenvalues (``wino-bino($+$)''). We perform a simulation with $4 \times 10^6$ events per parameter point prior to MLM merging, and simulate events for points in a uniform grid in the $(m_{\tilde{\chi}_2^0}, \Delta m)$ plane. The code chain is handled using \BSMArt \cite{Goodsell:2023iac}: batches of Les Houches event files are produced by gridpacks created by \madgraph using the same machinery as above, and then fed to \hackanalysis \cite{Goodsell:2021iwc,Goodsell:2024aig} which runs \pythia internally to perform merging and showering before conducting the analysis. This enables a large number of events to be simulated using a cluster in a relatively short space of time. The statistical analysis is then performed using routines included with \hackanalysis that employ \spey \cite{Araz:2023bwx} and \pyhf \cite{Heinrich:2021gyp}, for which we also require the next-to-leading-order (NLO) cross sections computed using {\sc Resummino}~\cite{Fuks:2013vua,Fiaschi:2023tkq}. The results are shown in the top plot of Figure~\ref{FIG:CMS2Llimits}. 

The agreement with the expected limits is good throughout the parameter space, and for the observed limit at small and large $\Delta m$. For moderate $\Delta m \in [5,20]$~GeV, however, the observed limit disagrees with that reported by CMS. It is highly curious that only the observed limit has a problem. We examined the signal yields and found that the limit is dominated in this region of the parameter space by contributions from two signal bins: namely the $2\ell$ Ultra-MET bin with $m_{\ell\ell} \in [4,10]$ GeV, and the $3\ell$ Med-MET bin. We show in in Table \ref{tab:SRMETbinsYields} the yields for the point $(m_{\tilde{\chi}_2^0}, \Delta m) = (280,10)$ GeV, which has a total cross section of $470$~fb, and the point $(m_{\tilde{\chi}_2^0}, \Delta m) = (320,10)$~GeV with total cross section of $280$~fb. In both spectra the second neutralino branching ratio to light leptons is $0.086$, and the table lists only bins for which we have non-negligible signal yields (compared to the background).
Essentially the limit is dominated therefore by just two bins where there is an underfluctuation. Given that the adjacent \met\ bins have \emph{over}fluctuations, it seems plausible that there is some overspill between them that would account for it in the measurements. In principle this should be taken into account in the statistical model by the correlations between the bins. However, this is clearly not enough using the simplified likelihood to yield the observed curve provided by CMS, whereas the expected limit \emph{is} well reproduced. 

\begin{table}
\centering
\captionsetup{width=0.85\textwidth}
\begin{tabular}{l|c|c|c|c} \toprule\toprule
  Bin name & Observed & Expected & Signal & Signal \\
    &            & background & $(280,10)$ & $(320,10)$ \\ \midrule
  $2\ell$ SR Med-MET $m_{\ell\ell} \in [4,10]$ & 19 & $17.8 \pm 4.4$ & $2.2$ & $1.4$\\
  $2\ell$ SR High-MET  $m_{\ell\ell} \in [4,10]$ & 11 & $7.7 \pm 3.9$ & $2.2$ & $1.3$ \\
  {\bf $2\ell$ SR Ultra-MET $m_{\ell\ell} \in [4,10]$} & 3 & $5.2 \pm 2.5$ & $5.2$ & $3.5$ \\
  $3\ell$ SR Low-MET  $m_{\ell\ell} \in [4,10]$ & 3 & $5.7 \pm 2.2$ & $0.7$ & $0.46$\\
   $3\ell$ SR Med-MET  $m_{\ell\ell} \in [1,4]$ & 3 & $1.7 \pm 1.1$ & $0.18 $ & $0.14$\\ 
 {\bf $3\ell$ SR Med-MET  $m_{\ell\ell} \in [4,10]$} & 1 & $4 \pm 1.8$ & $1.4$ & $1.0$\\ \bottomrule\bottomrule
\end{tabular}
\caption{\label{tab:SRMETbinsYields}Bins of the CMS-SUS-18-004 analysis with non-negligible yields compared to background for the benchmark points with $(m_{\tilde{\chi}^0_2},\Delta m) = (280,10)$~GeV and $(320,10)$~GeV.}
\end{table}


To investigate  this issue, we created a new simplified likelihood for this model by merging the ``Ultra'' and ``High'' $m_{\ell\ell} \in [4,10]$~GeV bins, and also by merging the $3\ell$ bins $m_{\ell\ell} \in [1,4]$~GeV and $m_{\ell\ell} \in [4,10]$~GeV. This is an entirely legitimate statistical procedure; in general we expect that it will weaken limits, but this is not necessarily the case, since aggregating bins together can allow several bins with small yields to give sufficient statistics to be relevant (indeed, that is why we have bins at all rather than treating each data point as its own bin). To do the combination, we aggregate bins as described in Eqs~\ref{eq:agreeg1} and \eqref{eq:agreeg2}. The motivation for these choices is twofold: it is likely that there is significant overspill between $m_{\ell\ell}$ bins at small invariant mass in the three-lepton case; and it is likely that there is either a mismodelling of the \met\ in our code or fluctuations in CMS' observations. Indeed, the bins with underfluctuations identified above are both adjacent to our selected bins with \emph{overfluctations} (compared to background), and it is possible that the simplified likelihood does not adequately take the correlations between them into account. We present limits for our exclusion plot in the bottom panel of Figure~\ref{FIG:CMS2Llimits}, where we see that the agreement with the CMS result is significantly improved. 

The fact that the difference between the observed limits is so stark throws into relief the limitations of the simplified likelihood in cases where there is an excess, especially where there are large numbers of signal regions. In addition, we are unable to adequately take the control regions into account. Alternatively, there could be a subtle error in our recast; but our recast agrees well with the results of CMS for the points provided, and in particular there is a point provided for the higgsino model, with $(m_{\tilde{\chi}_2^0}, \Delta m) = (180,10)$~GeV (which has the same $\Delta m$ as the region where the discrepancy between statistical models is greatest, albeit at smaller neutralino mass) where we see good agreement with the efficiencies and between the Ultra and High regions. The best way to resolve this would be to obtain a {\sc COMBINE} \cite{CMS:2024onh} model; equally, more efficiency data (concerning lepton and \met\ reconstruction) and information about additional signal points with larger $m_{\tilde{\chi}_2^0}$ would also be helpful.

Finally, as a byproduct of this investigation, we are also able to resolve the question of why the CMS analysis has better sensitivity compared to the equivalent ATLAS analysis around the region $\Delta m \sim 10$~GeV, when the CMS analysis contains a more straightforward set of cuts; ATLAS split into different regions for electron and muon pairs, and use {\sc RestFrames} \cite{restframes} variables to distinguish ISR jets: the simple distinction of the Ultra-MET bin yields a better discrimination of signal from background in that region, where the signal yields are small but the \met\ distribution of the background events falls off more rapidly.


\section{Recast of ATLAS-SUSY-2018-22 \texorpdfstring{($\boldsymbol{\geq 2}$ high-$\boldsymbol{p_{\text{T}}}$ jets, 0\,$\boldsymbol{\ell}$)}{2j+0l} in \texorpdfstring{\madanalysis}{MadAnalysis5}}
\label{sec:RecastMultijet}

The ATLAS multijet + \met\ search, ATLAS-SUSY-2018-22 \cite{ATLAS2020syg}, targets new physics in final states with at least two jets with high transverse momentum and no isolated energetic leptons. It uses 139~fb$^{-1}$ of LHC data at a centre-of-mass energy of 13~TeV, collected in the years 2015--2018 with the ATLAS detector. This analysis is sensitive to many possible supersymmetric (SUSY) scenarios, in particular when their signatures involve the production and decay of gluinos and squarks, and it is similar in terms of signals targeted to the CMS analysis CMS-SUS-19-006 \cite{CMS:2019zmd} which has already been implemented in \madanalysis \cite{DVN/4DEJQM_2020, Fuks:2021zbm}. A recast of this analysis in \checkmate \cite{Drees:2013wra,Dercks:2016npn,Desai:2021jsa} was employed in \cite{Buanes:2022wgm} to point out that it could be used for placing limits on electroweakinos. Hence having both the CMS and ATLAS multijet + \met\ searches implemented in \madanalysis will potentially allow us to probe our parameter space in regions where the monojet and soft-lepton searches are not sensitive. In \cite{Fuks:2024qdt} we exploited the available implementation of CMS-SUS-19-006 to constrain a model of \emph{frustrated dark matter} \cite{Carpenter:2022lhj}, and we compared the associated exclusions with limits originating from monojet searches. Our results showed that observed limits on scenarios featuring a small mass difference $\Delta m$ between the new physics states were similar, although the expected bounds were different; and that they were roughly independent of $\Delta m$, contrary to the monojet limits, which weaken with increasing $\Delta m$ values. We shall see here that this also holds true for electroweakino models, and we shall also (re-)examine different scenarios including the frustrated dark matter model of \cite{Fuks:2024qdt} by means of the new ATLAS-SUSY-2018-22 recast introduced in this section.

The ATLAS-SUSY-2018-22 analysis explores a search for SUSY particles through three distinct strategies focusing on final states composed exclusively of hadronic jets and significant missing transverse momentum. The first approach, known as the ``multi-bin search'', defines search regions that are mutually orthogonal, thus allowing for statistical combination to determine cross section upper limits. This is the strategy that we follow in this work. The second approach employs boosted decision trees for event selection, while the third approach, which is already implemented in the Public Analysis Database of \madanalysis\footnote{\url{https://dataverse.uclouvain.be/dataset.xhtml?persistentId=doi:10.14428/DVN/NW3NPG}}, is based on a traditional cut-and-count event selection method.
The multi-bin search extends previous ATLAS results using 36.1~fb$^{-1}$ of data \cite{ATLAS2017mjy} by simultaneously fitting the background expectations to the observed data yields in multiple event selection bin, hence leading to enhanced sensitivity. Moreover, the ATLAS collaboration made available substantial additional data via \textsc{HepData}\footnote{\url{https://www.hepdata.net/record/95664}}, including detailed cutflow tables and exclusion curves as well as digitised information on the figures useful for the validation of any recast implementation. 

\subsection{Description of the analysis}

The ATLAS-SUSY-2018-22 analysis is based on the selection of events that exhibit a significant momentum imbalance in the transverse plane (\emph{i.e.}\ with large $\met$). Isolated leptons with pseudorapidity $|\eta| < 2.7$ (2.47) are vetoed if their $p_{\text{T}} > 6~(7)$~GeV for muons (electrons), respectively, and signal jets are required to have $p_{\text{T}} > 50$~GeV and pseudorapidity $|\eta| < 2.8$. A significant amount of hadronic activity is expected, characterised by a high number of jets $N_j$ and large (exclusive) hadronic transverse energy $H_{\text{T}}$ defined as the scalar sum of the transverse momenta of all jets with $p_{\text{T}} > 50~\text{GeV}$ and $|\eta| < 2.8$. Additionally, events must exhibit large values of the ``effective mass" variable defined as $m_{\text{eff}} = H_{\text{T}} + E_{\text{T}}^{\text{miss}}$, and the missing transverse momentum must be well separated from any jet candidate. All of these criteria are summarised in Table~\ref{tab:ATLAS_preselection}.

\renewcommand{\arraystretch}{1.1}
\begin{table}
\captionsetup{width=0.85\textwidth}
\centering
\begin{tabular}{lc}
\toprule\toprule
Lepton veto & No baseline electron (muon) with $p_{\text{T}} > 7\,(6)$ GeV \\
$E_{\text{T}}^{\text{miss}}$ [GeV] & $> 300$ \\
$p_{\text{T}}(j_1)$ [GeV] & $> 200$ \\
$p_{\text{T}}(j_2)$ [GeV] & $> 50$ \\
$\Delta\phi(j_{1,2,(3)},\vec{p}_{\text{T}}^{\text{miss}})_{\text{min}}$ & $> 0.2$ \\
$m_{\text{eff}}$ [GeV] & $> 800$ \\
\bottomrule\bottomrule
\end{tabular}
\caption{\label{tab:ATLAS_preselection}Summary of common preselection criteria used for the multi-bin selection strategy of the ATLAS-SUSY-2018-22 search.}
\end{table}
\renewcommand{\arraystretch}{1.0}
 
Squarks typically produce at least one jet in their decays, for instance through a $\tilde{q} \rightarrow q \tilde{\chi}_1^0$ decay, while gluinos typically produce at least two jets, for instance through a $\tilde{g} \rightarrow qq \tilde{\chi}_1^0$ decay. The production of $\tilde{q} \tilde{q}^*$ and $\tilde{g} \tilde{g}$ pairs therefore leads to events containing at least two or four jets, respectively, in addition to missing transverse energy carried away by the neutralinos. Moreover, the decays of heavy SUSY and SM particles (like hadronically-decaying $W$ or $Z$ bosons) produced in longer $\tilde{q}$ and $\tilde{g}$ decay cascades tend to further increase the jet multiplicity in the final state.  
To target different SUSY particle production and decay scenarios, signal regions with varying jet multiplicity requirements and specific ranges of kinematic variables are defined in the multi-bin search. It thus includes three sets of signal regions: the MB-SSd (‘multi-bin squark-squark-direct’) regions target scenarios with a large mass difference between pair-produced squarks and the $\tilde{\chi}_1^0$, the MB-GGd (‘multi-bin gluino-gluino-direct’) regions target scenarios with a large mass difference between pair-produced gluinos and the $\tilde{\chi}_1^0$, and the MB-C (‘multi-bin compressed’) regions target scenarios with a small mass difference between pair-produced squarks or gluinos and the $\tilde{\chi}_1^0$.

In practice, events are assigned to one of these three classes of mutually exclusive signal regions based on the jet multiplicity, the effective mass $m_{\text{eff}}$ and the missing transverse momentum significance defined as $E_{\text{T}}^{\text{miss}} / \sqrt{H_{\text{T}}}$. This variable is used to suppress backgrounds where jet energy mismeasurement generates missing transverse momentum, and was found to enhance sensitivity to signals emerging from $\tilde{q} \tilde{q}^*$ production. To define the various multi-bin search regions, at least two jets with $|\eta| < 2$ are required for the MB-SSd regions, with the sub-leading jet having $p_{\text{T}} > 100~\text{GeV}$. The MB-C regions rely instead on selecting a single energetic jet with $p_{\text{T}} > 600~\text{GeV}$, while in the MB-GGd regions, at least four jets with $p_{\text{T}} > 100~\text{GeV}$ and $|\eta| < 2$ are required. The azimuthal separation between the missing transverse momentum and the three leading jets $\Delta\phi_{\text{min}}(j_{1,2,3}, \vec{p}_{\text{T}}^{\text{miss}})$ must exceed 0.4, while for other jets with $p_{\text{T}} > 50~\text{GeV}$, $\Delta\phi_{\text{min}}(j_i, \vec{p}_{\text{T}}^{\text{miss}})$ must exceed 0.2. For the MB-SSd regions, tighter thresholds of 0.8 and 0.4 are applied to reduce the multijet background caused by jet energy mismeasurement, and the aplanarity variable $A = \frac{3}{2}\lambda_3$, with $\lambda_3$ being the smallest eigenvalue of the normalised momentum tensor, is additionally used to enhance sensitivity to signal processes, with a requirement of $A > 0.04$. Finally, the missing transverse momentum significance $E_{\text{T}}^{\text{miss}}/\sqrt{H_{\text{T}}}$ is required to be greater than 10~GeV$^{1/2}$ in all regions, and we further impose $m_{\text{eff}} > 1000~\text{GeV}$ except in the MB-C regions where a tighter condition $m_{\text{eff}} > 1600~\text{GeV}$ is applied. All these conditions are summarised in Table~\ref{tab:ATLAS_selection}. 

\renewcommand{\arraystretch}{1.1}
\begin{table}
\captionsetup{width=0.85\textwidth}
\centering
\begin{tabular}{lccc}
\toprule\toprule
& MB-SSd & MB-GGd & MB-C \\
\midrule
$N_j$ & $\geq 2$ & $\geq 4$ & $\geq 2$ \\
$p_{\text{T}}(j_1)$ [GeV] & $> 200$ & $> 200$ & $> 600$ \\
$p_{\text{T}}(j_{i=2,\ldots,N_{\text{jmin}}})$ [GeV] & $> 100$ & $> 100$ & $> 50$ \\
$|\eta(j_{i=1,\ldots,N_{\text{jmin}}})|$ & $< 2.0$ & $< 2.0$ & $< 2.8$ \\
$\Delta\phi(j_{1,2,(3)},\vec{p}_{\text{T}}^{\text{miss}})_{\text{min}}$ & $> 0.8$ & $> 0.4$ & $> 0.4$ \\
$\Delta\phi(j_{i>3},\vec{p}_{\text{T}}^{\text{miss}})_{\text{min}}$ & $> 0.4$ & $> 0.4$ & $> 0.2$ \\
Aplanarity & $-$ & $> 0.04$ & $-$ \\
$E_{\text{T}}^{\text{miss}}/\sqrt{H_{\text{T}}}$ [GeV$^{1/2}$] & $> 10$ & $> 10$ & $> 10$ \\
$m_{\text{eff}}$ [GeV] & $> 1000$ & $> 1000$ & $> 1600$ \\
\bottomrule\bottomrule
\end{tabular}
\caption{\label{tab:ATLAS_selection}Summary of selection criteria used for the multi-bin search.}
\end{table}
\renewcommand{\arraystretch}{1.0}

Additionally, the three sets of signal regions are further binned in terms of the variables $m_{\text{eff}}$, $E_{\text{T}}^{\text{miss}}/\sqrt{H_{\text{T}}}$ and the number of jets $N_j$. For the MB-SSd regions, two jet multiplicity bins, six $m_{\text{eff}}$ bins and four $E_{\text{T}}^{\text{miss}}/\sqrt{H_{\text{T}}}$ bins are considered, which results in 24 MB-SSd signal regions. In the low jet multiplicity bin ($N_j = 2$ or $3$), tighter requirements are applied to the leading and sub-leading jets such that $p_{\text{T}}(j_{1,2}) > 250~\text{GeV}$. In addition, some bins are merged to reduce the total number of signal regions without significant loss of sensitivity. For the MB-GGd regions, six $m_{\text{eff}}$ bins and three $E_{\text{T}}^{\text{miss}}/\sqrt{H_{\text{T}}}$ bins are introduced; for the MB-C regions, three jet multiplicity bins, three $m_{\text{eff}}$ bins and two $E_{\text{T}}^{\text{miss}}/\sqrt{H_{\text{T}}}$ bins are instead considered.

\subsection{Validation}

For the purpose of the validation, we focus on the MB-GGd regions, but similar results can be obtained for the MB-C and the MB-SSd regions. Signal events are generated using the same toolchain as described in Section~\ref{sec:RecastSoftLeptons}. To match the cutflows provided by the ATLAS collaboration, $5\times10^5$ events are simulated at leading order, and signal cross sections are calculated to approximate next-to-next-to-leading order (NNLO) in the strong coupling constant, including the resummation of soft-gluon emission at next-to-next-to-leading-logarithmic (NNLL) accuracy using \textsc{NNLL-fast} v2.0~\cite{Beenakker:2016lwe, Beenakker:2024jwh}. Relying on the material provided in the \textsc{HepData} repository of the ATLAS-SUSY-2018-22 analysis, we compare in Table~\ref{tab:mean_cross_section_comparison} our predictions to the ATLAS official results for various scenarios defined in terms of the gluino and neutralino masses, estimating the difference relative to the official ATLAS estimates.

\begin{table}
\captionsetup{width=0.85\textwidth}\setlength\arraycolsep{4pt}\renewcommand{\arraystretch}{1.25}
\centering
\resizebox{0.48\textwidth}{!}{\begin{tabular}{l|c c S[table-format=3.2]}
\toprule\toprule
Bin name & ATLAS & Recast & \text{Difference\,(\%)} \\ \midrule
SR\_4\_1000\_10\_cuts  & $8.63\times 10^{-3}$ & $1.20\times 10^{-2}$ & -39.27 \\  
SR\_4\_1000\_16\_cuts  & $5.58\times 10^{-3}$ & $5.60\times 10^{-3}$ & -0.39  \\  
SR\_4\_1000\_22\_cuts  & $7.10\times 10^{-4}$ & $9.08\times 10^{-4}$ & -27.83 \\  
SR\_4\_1600\_10\_cuts  & $4.11\times 10^{-2}$ & $4.12\times 10^{-2}$ & -0.12  \\  
SR\_4\_1600\_16\_cuts  & $3.83\times 10^{-2}$ & $3.81\times 10^{-2}$ & 0.67   \\  
SR\_4\_1600\_22\_cuts  & $1.81\times 10^{-2}$ & $2.07\times 10^{-2}$ & -14.12 \\  
SR\_4\_2200\_10\_cuts  & $2.34\times 10^{-2}$ & $1.99\times 10^{-2}$ & 15.26  \\  
SR\_4\_2200\_16\_cuts  & $3.09\times 10^{-2}$ & $2.76\times 10^{-2}$ & 10.80  \\  
SR\_4\_2200\_22\_cuts  & $4.30\times 10^{-2}$ & $4.01\times 10^{-2}$ & 6.77   \\  
SR\_4\_2800\_10\_cuts  & $5.56\times 10^{-3}$ & $4.16\times 10^{-3}$ & 25.06  \\  
SR\_4\_2800\_16\_cuts  & $8.16\times 10^{-3}$ & $6.60\times 10^{-3}$ & 19.17  \\  
SR\_4\_2800\_22\_cuts  & $1.43\times 10^{-2}$ & $1.17\times 10^{-2}$ & 18.53  \\  
SR\_4\_3400\_10\_cuts  & $1.01\times 10^{-3}$ & $8.19\times 10^{-4}$ & 19.23  \\  
SR\_4\_3400\_16\_cuts  & $3.54\times 10^{-4}$ & $1.02\times 10^{-3}$ & -187.77 \\  
SR\_4\_3400\_22\_cuts  & $1.75\times 10^{-3}$ & $2.46\times 10^{-3}$ & -40.68 \\  
SR\_4\_4000\_16\_cuts  & $2.59\times 10^{-4}$ & $1.33\times 10^{-4}$ & 48.62  \\  
SR\_4\_4000\_22\_cuts  & $9.92\times 10^{-4}$ & $2.66\times 10^{-4}$ & 73.22  \\  \bottomrule\bottomrule
\end{tabular}}\hfill 
\resizebox{0.48\textwidth}{!}{\begin{tabular}{l|c c S[table-format=3.2]}
\toprule\toprule
Bin name & ATLAS & Recast & \text{Difference\,(\%)} \\ \midrule
SR\_4\_1000\_10\_cuts  & $8.72\times 10^{-3}$ & $7.89\times 10^{-3}$ & 9.57   \\  
SR\_4\_1000\_16\_cuts  & $1.88\times 10^{-3}$ & $2.80\times 10^{-3}$ & -49.03 \\  
SR\_4\_1000\_22\_cuts  & $2.86\times 10^{-4}$ & $4.21\times 10^{-4}$ & -47.04 \\  
SR\_4\_1600\_10\_cuts  & $5.03\times 10^{-2}$ & $3.54\times 10^{-2}$ & 29.70  \\  
SR\_4\_1600\_16\_cuts  & $4.26\times 10^{-2}$ & $2.83\times 10^{-2}$ & 33.60  \\  
SR\_4\_1600\_22\_cuts  & $1.30\times 10^{-2}$ & $1.33\times 10^{-2}$ & -2.71  \\  
SR\_4\_2200\_10\_cuts  & $3.79\times 10^{-2}$ & $2.77\times 10^{-2}$ & 26.74  \\  
SR\_4\_2200\_16\_cuts  & $4.89\times 10^{-2}$ & $3.31\times 10^{-2}$ & 32.37  \\  
SR\_4\_2200\_22\_cuts  & $4.05\times 10^{-2}$ & $3.68\times 10^{-2}$ & 9.08   \\  
SR\_4\_2800\_10\_cuts  & $7.06\times 10^{-3}$ & $7.09\times 10^{-3}$ & -0.45  \\  
SR\_4\_2800\_16\_cuts  & $1.08\times 10^{-2}$ & $9.72\times 10^{-3}$ & 9.95   \\  
SR\_4\_2800\_22\_cuts  & $1.57\times 10^{-2}$ & $1.36\times 10^{-2}$ & 13.05  \\  
SR\_4\_3400\_10\_cuts  & $1.65\times 10^{-3}$ & $1.24\times 10^{-3}$ & 24.66  \\  
SR\_4\_3400\_16\_cuts  & $8.96\times 10^{-4}$ & $1.48\times 10^{-3}$ & -65.57 \\  
SR\_4\_3400\_22\_cuts  & $4.42\times 10^{-3}$ & $3.08\times 10^{-3}$ & 30.35  \\  
SR\_4\_4000\_10\_cuts  & $7.10\times 10^{-4}$ & $2.11\times 10^{-4}$ & 70.22  \\  
SR\_4\_4000\_16\_cuts  & $3.12\times 10^{-4}$ & $4.23\times 10^{-4}$ & -35.40 \\  
SR\_4\_4000\_22\_cuts  & $3.02\times 10^{-4}$ & $6.57\times 10^{-4}$ & -117.47 \\ \bottomrule\bottomrule
\end{tabular}}\vspace{.8cm}
\resizebox{0.48\textwidth}{!}{\begin{tabular}{l|c c S[table-format=3.2]}
\toprule\toprule
Bin name & ATLAS & Recast & \text{Difference\,(\%)} \\ \midrule
SR\_4\_1000\_10\_cuts  & $4.27\times 10^{-2}$ & $4.67\times 10^{-2}$ & -9.21  \\  
SR\_4\_1000\_16\_cuts  & $1.72\times 10^{-2}$ & $2.94\times 10^{-2}$ & -70.93 \\  
SR\_4\_1000\_22\_cuts  & $1.99\times 10^{-3}$ & $4.32\times 10^{-3}$ & -117.65 \\  
SR\_4\_1600\_10\_cuts  & $1.24\times 10^{-2}$ & $8.63\times 10^{-3}$ & 30.23  \\  
SR\_4\_1600\_16\_cuts  & $8.54\times 10^{-3}$ & $7.25\times 10^{-3}$ & 15.01  \\  
SR\_4\_1600\_22\_cuts  & $2.61\times 10^{-3}$ & $2.93\times 10^{-3}$ & -12.30 \\  
SR\_4\_2200\_10\_cuts  & $1.35\times 10^{-3}$ & $1.24\times 10^{-3}$ & 8.32   \\  
SR\_4\_2200\_16\_cuts  & $1.45\times 10^{-3}$ & $1.49\times 10^{-3}$ & -2.61  \\  
SR\_4\_2200\_22\_cuts  & $6.67\times 10^{-4}$ & $1.03\times 10^{-3}$ & -53.80 \\  
SR\_4\_2800\_10\_cuts  & $1.99\times 10^{-4}$ & $2.28\times 10^{-4}$ & -14.47 \\  
SR\_4\_2800\_16\_cuts  & $2.99\times 10^{-4}$ & $2.39\times 10^{-4}$ & 20.10  \\  
SR\_4\_2800\_22\_cuts  & $1.85\times 10^{-4}$ & $3.47\times 10^{-4}$ & -87.35 \\ \bottomrule\bottomrule
\end{tabular}}
\caption{\label{tab:mean_cross_section_comparison}MG-GGd selection efficiencies as provided by ATLAS and using our \madanalysis\ implementation, the differences being given relative to the ATLAS results. We consider scenarios with $(m_{\tilde{g}}, m_{\tilde{\chi}_1^0}) = (2200, 1300)$~GeV (top left), (1800, 800)~GeV (top right) and (1300, 900)~GeV (bottom) with respective signal NNLO+NNLL rates of 0.357, 2.94 and 52.3~fb$^{-1}$.}
\end{table}

\begin{figure}
\captionsetup{width=0.85\textwidth}
\centering
\includegraphics[width=0.8\textwidth]{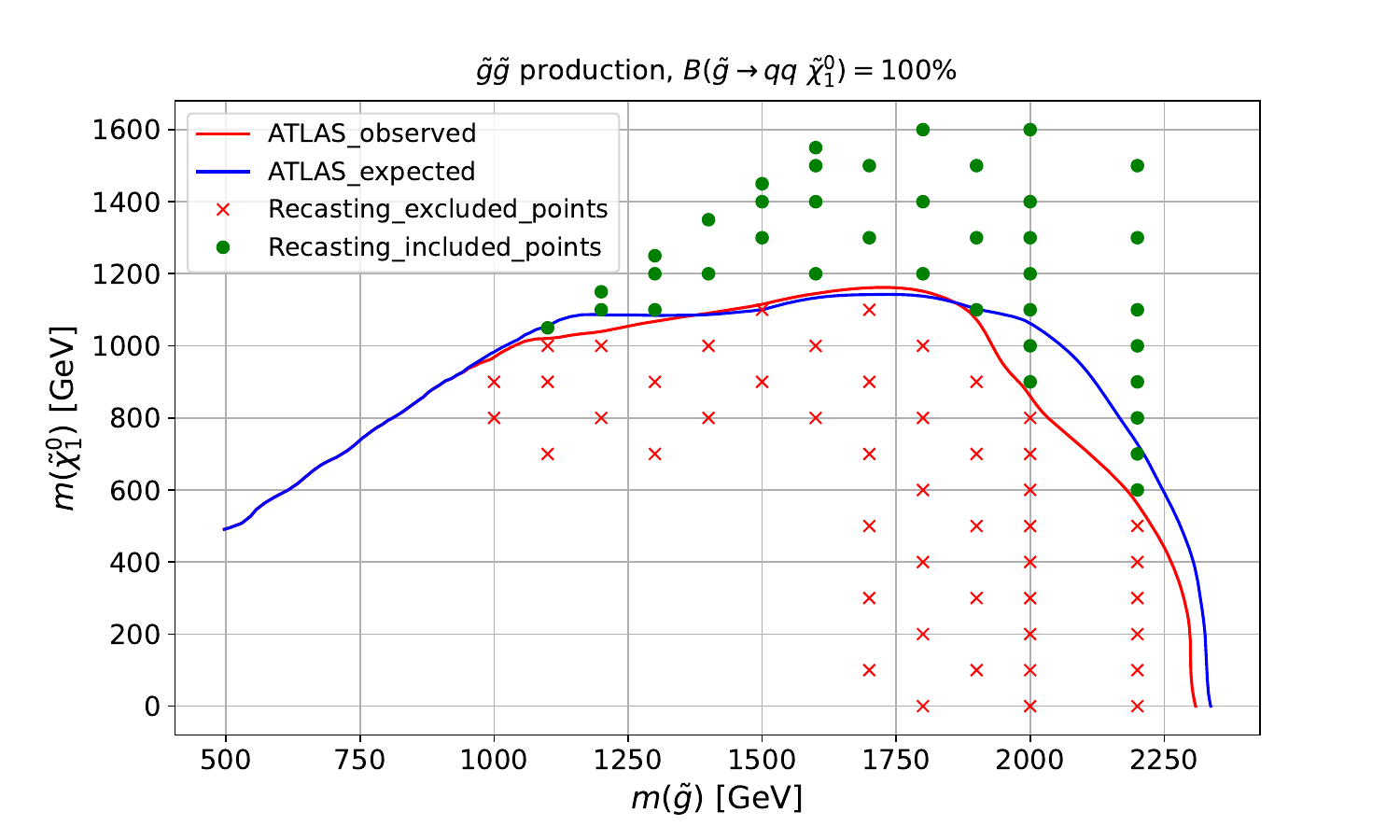} 
\caption[width=\textwidth]{\label{fig:ATLAS_validation}95\% confidence level exclusion contours in the $(m_{\tilde{g}}, m_{\tilde{\chi}_1^0})$ mass plane. We compare  the observed (red) and expected (blue) limits provided by ATLAS to predictions obtained with our implementation of ATLAS-SUSY-2018-22 in \madanalysis for gluino pair production followed by the decay $\tilde{g} \rightarrow q \bar{q} \tilde{\chi}^0_1$. Red cross markers represent excluded points while green dot markers correspond to allowed points.}
\end{figure}

We achieve good agreement with the ATLAS predictions overall, except for a few bins, likely due to cross section miscalculations or uncertainty effects. Since these discrepancies do not consistently occur in the same bins for each mass point, they are not easy to investigate without more information from the ATLAS collaboration. Consequently, we have decided to produce an 95\% confidence level exclusion contour in the gluino-neutralino mass plane to further confirm the validity of our recast, considering gluino pair production followed by the decay $\tilde{g} \rightarrow q \bar{q} \tilde{\chi}^0_1$. The results are presented in Figure~\ref{fig:ATLAS_validation} in which we superimpose the official expected (blue) and observed (red) contours from ATLAS and the results expected from our recast implementation in \madanalysis. Green dots correspond to allowed mass configurations while red crosses denote excluded mass configurations. As can be seen, an excellent degree of agreement is found despite the aforementioned discrepancies.

\subsection{Applications}

As mentioned above, analyses of multijet + \met\ final states at the LHC offer valuable sensitivity to compressed electroweakino scenarios since their signal(s) can comprise missing transverse energy produced in association with hard jets. In particular, the CMS Run 2 multijet + \met\ search (CMS-SUS-19-006) \cite{CMS:2019zmd} has proven useful for constraining such models, especially when combined with monojet results as demonstrated in \cite{Fuks:2024qdt}. Although multijet + \met\ searches typically yield only mild exclusions for simplified higgsino models, they can lead to significantly stronger constraints on wino-bino scenarios. Furthermore, multijet+\met\  analyses are also particularly relevant for testing non-minimal models such as the \textit{compressed frustrated dark matter} (CFDM) scenario previously introduced in \cite{Fuks:2024qdt} within the framework developed in \cite{Carpenter:2022lhj}. In this section, we present updated results for those models from our new ATLAS-SUSY-2018-22 analysis implementation in \madanalysis. 

For the higgsino and wino-bino setups, the derived exclusion limits are generally weaker than those obtained from existing monojet and soft-lepton analyses. For instance, the new constraint on the higgsino parameter space exhibits a vertical cutoff near $m_{\tilde{\chi}^0_1} \approx 120$~GeV. Similarly, for the wino-bino model, weaker exclusions are obtained compared to those derived from the CMS soft-lepton search, which will be analysed in detail in Section~\ref{sec:LimitInterpretation}. Nonetheless, these results, summarised in Table~\ref{tab:multijet_constraints}, illustrate the complementary role that multijet + \met\ analyses can in principle play in probing electroweakino models, particularly when combined with other LHC searches, as will be shown in the next section.

\begin{table}
\captionsetup{width=0.85\textwidth}
\centering
\caption{\label{tab:multijet_constraints} 95\%  confidence level exclusion constraints derived with the reinterpretation of the results of the ATLAS-SUSY-2018-22 multijet + \met\ analysis for two sets of simplified electroweakino models.}
\begin{tabular}{lc}
\toprule
\toprule
Model & Excluded region (approx.) \\
\midrule
Simplified higgsino & $m_{\tilde{\chi}^0_1} \lesssim 120$~GeV \\
Simplified wino-bino & $m_{\tilde{\chi}^0_1} \lesssim 170$~GeV \\
\bottomrule\bottomrule
\end{tabular}
\end{table}

\begin{figure}
\captionsetup{width=0.85\textwidth}
\centering
\includegraphics[width=0.8\textwidth]{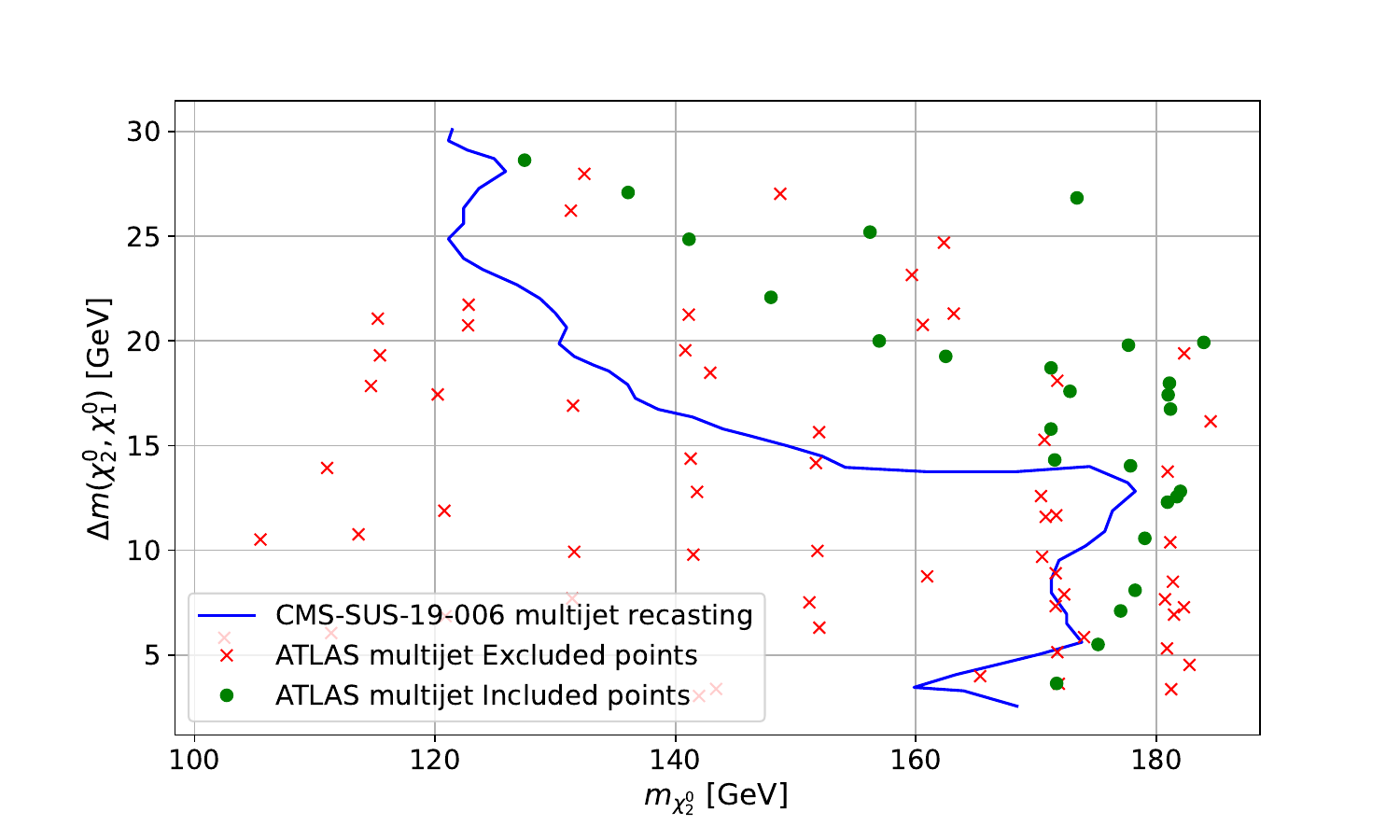} 
\caption{\label{fig:SMTR_Multijet}95\%  confidence level exclusion constraints on the CFDM model obtained from the CMS-SUS-19-006 (blue) and ATLAS-SUSY-2018-22 (red crosses and green dots) multijet+\met\ analyses.}
\end{figure}

Finally, we also examine the CFDM model. This scenario extends the hyperchargeless Higgs triplet model by incorporating additional electroweak fermions in a frustrated dark sector configuration. A distinctive feature of this model is the loop-induced coupling of a neutral scalar to diphotons, which has been proposed as a potential explanation for the observed diphoton excess around 152~GeV \cite{Ashanujjaman:2024pky,Crivellin:2024uhc,Banik:2024ftv}. Additionally, the model predicts an enhanced monojet production cross section, and subsequentially potentially significant multijet+\met\ production. The constraints derived for this model are shown in Figure~\ref{fig:SMTR_Multijet}. The ATLAS bounds, shown through red crosses (excluded configurations) and green dots (allowed configurations) are slightly stronger than those obtained from the corresponding CMS search shown through the blue contour. This once again motivates a potential combination of the bounds, which we leave for future work.

\section{Limits and preferred parameter space for the wino-bino model}
\label{sec:LimitInterpretation}

\subsection{Statistical presentation}

In the following we shall present limits on the wino-bino(+) simplified model, including those from CMS monojet searches, and a combination of all monojet and soft-lepton searches considered in this study. We shall not include the CMS or ATLAS multijet searches since, as noted above, they are not sensitive to the interesting portion of the parameter space, and it is not clear to what extent they are statistically independent of the monojet searches. However, since we are interested in the excesses in these searches, we shall also attempt to quantify these excesses in different regions of the parameter space, for each search individually and combined. In order to do this, we follow the approach advocated in \cite{Fowlie:2024dgj} and used by us in \cite{Goodsell:2024aig,Fuks:2024qdt} to plot contours of the ratio of likelihoods of the signal + background hypothesis compared to background-only hypothesis. By abuse of notation we refer to this quantity as the ``Bayes factor'' and denote it $B_{10}$, defined as
\begin{align}
B_{10} \equiv \frac{\mathcal{L}(1, \hat{\hat{\theta}}(1))}{\mathcal{L}(0,\hat{\hat{\theta}}(0))}.
\end{align}
where $\mathcal{L}(\mu, \theta)$ is the likelihood of the data given signal strength $\mu$ and nuisance parameters $\theta;$ and $\hat{\hat{\theta}}(\mu)$ means the values of those nuisance parameters that maximise the likelihood for the given, fixed, signal strength. For the signal + background model, $\mu=1$, and for the background $\mu = 0$.

The profiling of the likelihood to obtain maxima is what is usually obtainable from the tools \spey and \pyhf that we are using, so this is particularly convenient to calculate. Moreover, it strikingly shows the compatibility -- or otherwise -- of different regions of the parameter space with the data, and the preference compared to the Standard Model (\emph{i.e.}\ the background). On the other hand, it does not have a strict statistical interpretation; ideally, we would be able to compute the ratio of Bayesian evidences instead (hence our calling $B_{10}$ the ``Bayes factor'', even if this is not rigorous), but this is not currently possible, and it is hoped that the above ratio is a good approximation (in this we emulate \cite{Fowlie:2024dgj}). It also has the advantage over traditional measures, such as the $p$-value, in that it compares the agreement of a \emph{specific} signal strength instead of the \emph{best-fit} signal strength. This is helpful for our purposes because in order to construct a traditional likelihood ratio it is necessary to have a complete set of hypotheses, which means allowing the signal strength to vary. But our models have well predicted signal strengths that we cannot vary. Hence we choose to plot contours of $B_{10}$ and identify the maxima, and list $p$-values for the best-fit points along with the best-fit signal strength (which ought to be close to $1$ for these cases).

A simple illustration of the difference between the statistical measures can be seen by considering a single-bin model without any uncertainty. Suppose that the expected background is Poisson-distributed with mean $n_{\text{b}}$ and the observed number of events $n_{\rm obs},$ while the predicted number of signal events is $n_{\text{s}}$. Computing the $p$-value is \emph{independent of $n_{\text{s}}$} in this case, because the signal strength is allowed to vary in the computation. The $p$-value in this idealised example therefore tells us nothing about the underlying model! However, the ratio of likelihoods 
$$
B_{10} \!\!\underset{\begin{subarray}{c}
  \rm toy\\ \rm single\\ \rm bin\end{subarray}}{=}\! \exp(-n_{\text{s}}) \left( 1 + \frac{n_{\text{s}}}{n_{\text{b}}} \right)^{n_{\rm obs}}
$$
has a maximum when $n_{\text{s}} = \mathrm{max}(0,n_{\rm obs} - n_{\text{b}})$ for given values of $n_{\text{b}}, n_{\rm obs}$, so the value of $B_{10}$ can  take any value from zero (if $n_{\text{s}}$ is too large compared to $n_{\rm obs}$) to infinity (if $n_{\rm obs} \gg n_{\text{b}}$). Hence these two quantities are complementary: $B_{10}$ tells us where the data is best fit, and the $p-$value roughly gives us the probability that the data represent a fluctuation of the background.
 
\subsection{Results}

\begin{figure}
\captionsetup{width=0.85\textwidth}
\centering
  \includegraphics[width=0.485\textwidth]{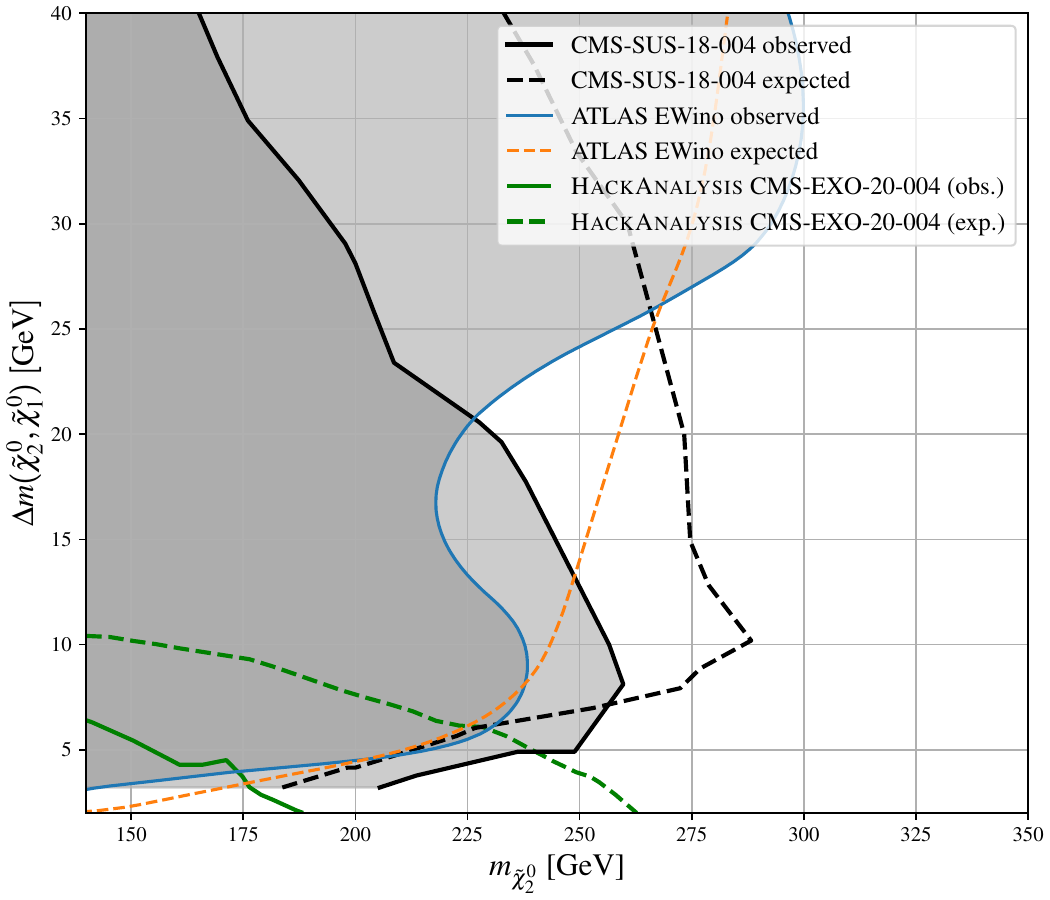}\hfill
  \includegraphics[width=0.485\textwidth]{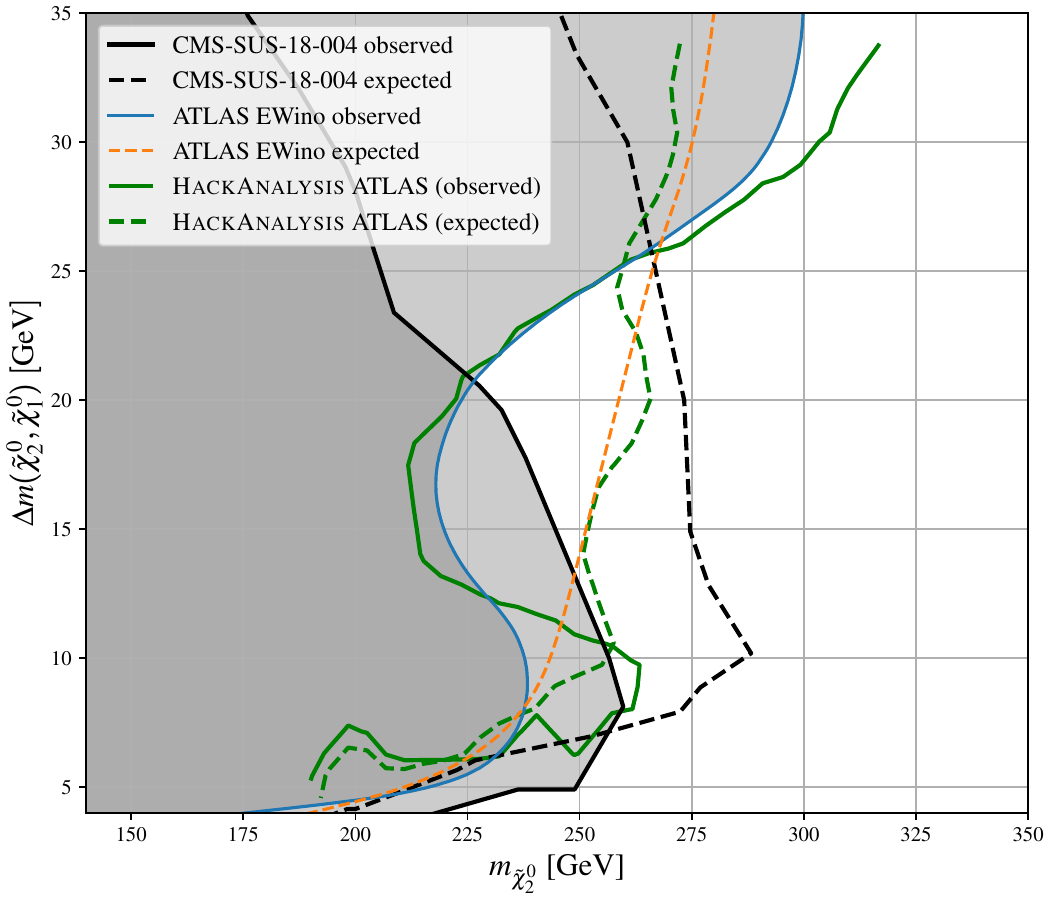}\\[1cm]
 \includegraphics[width=0.485\textwidth]{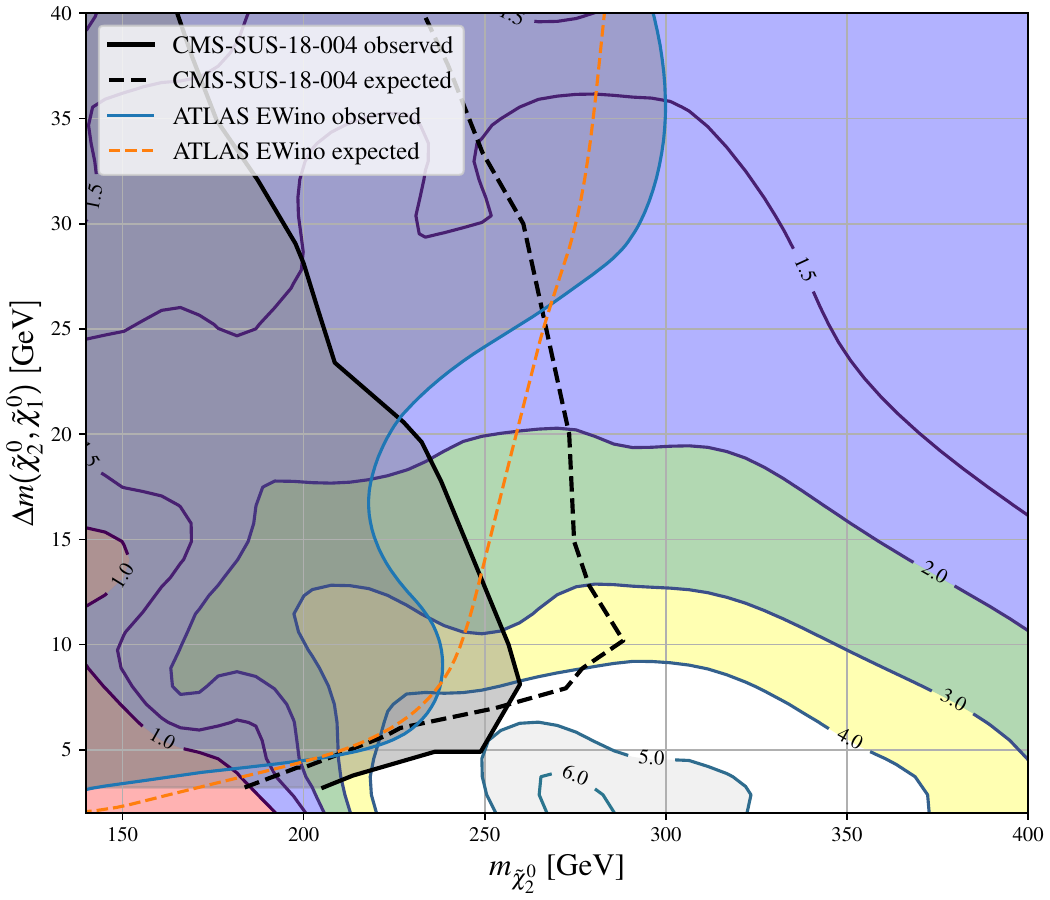} \hfill \includegraphics[width=0.485\textwidth]{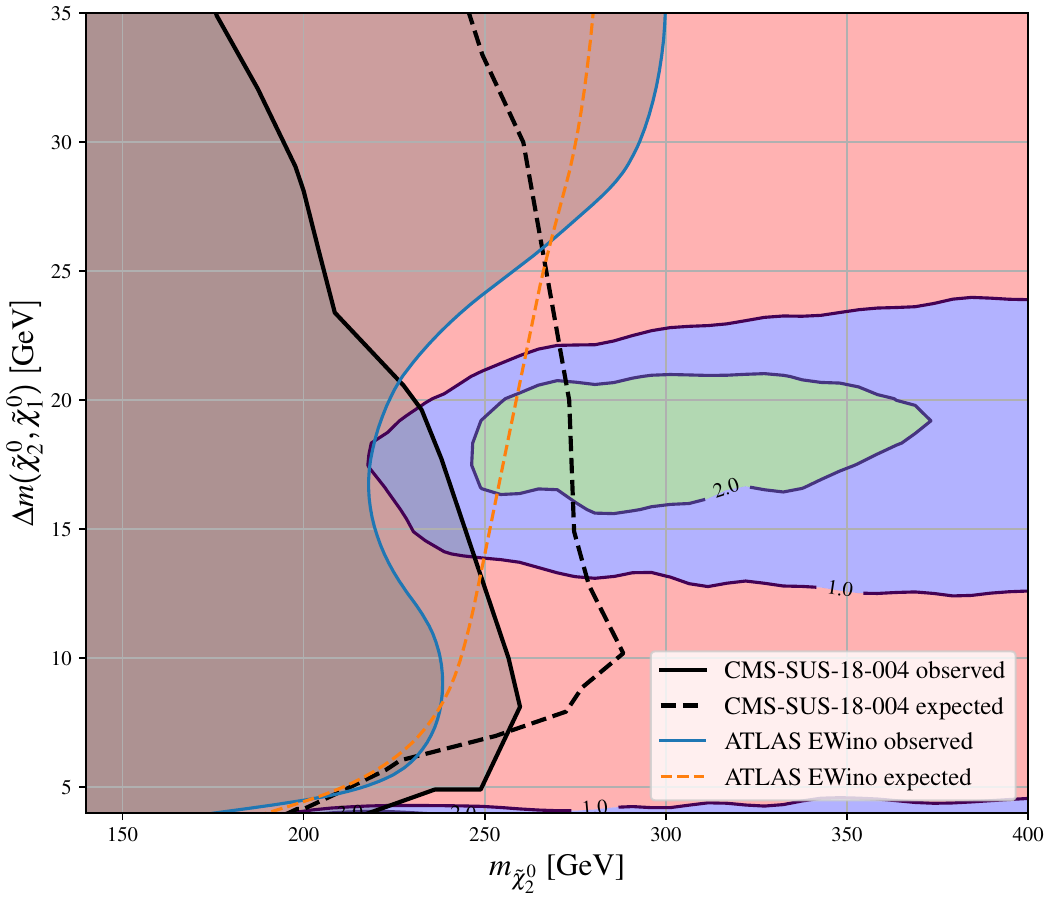} 
  \caption{\label{FIG:Monojet}Results for CMS monojet searches (left) and ATLAS soft-lepton+\met\ (right) in the wino-bino scenario, overlaid with soft-lepton excluded regions reported by ATLAS and CMS. Top: Expected and observed limits. Bottom: Contours of Bayes factor $B_{10}$.}
\end{figure}

We now present the application of our set of recasts to the simplified wino-bino scenario. The samples are generated using the same toolchain as for the validation in Section~\ref{sec:CMSValidation}; we have also made available all the inputs and results from this section in a  {\sc Zenodo} dataset \cite{fuks_2025_15727335}\footnote{\url{https://zenodo.org/records/15727335}}. Firstly, in the top left panel of Figure~\ref{FIG:Monojet} we show the limits arising from the CMS monojet search only. The difference between observed and expected limits is clearly seen. This plot equivalent to the limits shown in \cite{Agin:2023yoq} for the higgsino scenario; it also confirms that the monojet search is much more sensitive than the multijet search at low $\Delta m,$ as can be seen by comparing the plot to the results in Table~\ref{tab:multijet_constraints}. The multijet limits essentially rule out a region of the parameter space already excluded by soft-lepton and monojet searches, so we will not consider them in the following. To identify the region of parameter space preferred by the monojet search, where there is an excess, we also show, in the bottom left panel of the figure, the results for our Bayes factor contours for the first time. These show that monojet results prefer a region with $m_{\tilde{\chi}_2^0} \in [250,300]$ GeV and small mass splittings, and that a very large area of the parameter space fits the data better than the SM, \emph{i.e.} with $B_{10} > 1$. This latter statement means that essentially \emph{any} scenario predicting the production of monojet events improves the fit with the data, and only the region with too many monojet events is disfavoured; but as the mass of the neutralino and the mass difference between the electroweakinos increase, the number of such events reduces to zero. Hence we show the $B_{10} = 1.5$ contour to show where this preference becomes small. This plot corroborates the claim in \cite{Agin:2023yoq} that the monojet excesses can come from the same models that generate soft-lepton excesses.

We show the Bayes contours for just the ATLAS soft-lepton searches in the right panel of Figure~\ref{FIG:Monojet}. In the top right plot, we show the limits as computed in \HackAnalysis compared to those given by ATLAS. These can be compared to the validation plots in \cite{Agin:2024yfs}: as compared to the earlier paper, we have added the control regions, but \emph{we do not include the signal uncertainty used by ATLAS}. The control regions improve the fit even compared to the agreement in \cite{Agin:2024yfs}; and the lack of signal uncertainty accounts for the small difference around $\Delta m \simeq 10$ GeV. The reason is that the signal regions with largest signal at small $\Delta m$ contain very few background events, so the uncertainty has a large effect; at larger $\Delta m$ there is larger background and the curves agree very well for both observed and expected limits. The Bayes factor contours in the bottom right panel of the figure are novel and show the preferred region. It is clear that only a relatively modest excess is manifest in this model (in contrast, using a ``generalised higgsino template'' ATLAS found an excess of more than $3\sigma$ \cite{ATLAS:2019lng,ATLAS:2021moa}); but, as we will see, the other excesses are compatible with this one and turn out to be in a phenomenologically interesting region of wino-bino parameter space.

\begin{figure}
\captionsetup{width=0.85\textwidth}
\centering
  \includegraphics[width=0.485\textwidth]{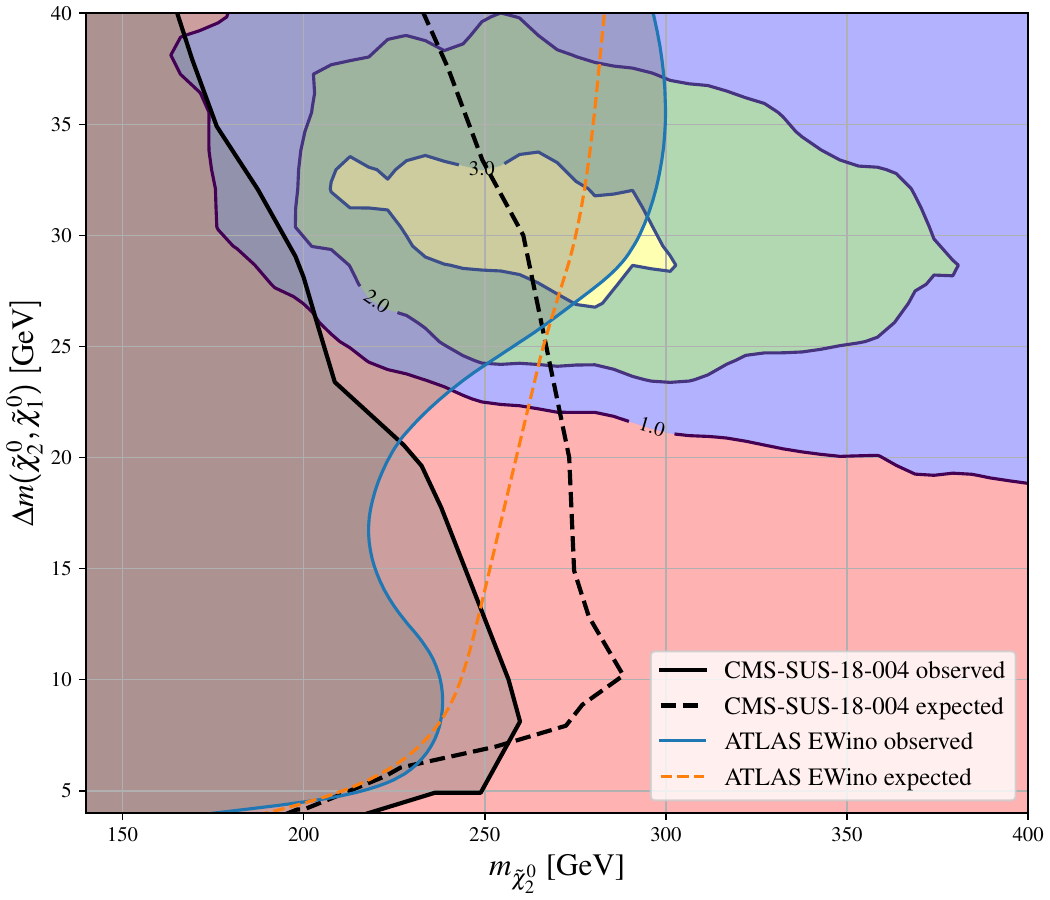}\hfill 
  \includegraphics[width=0.485\textwidth]{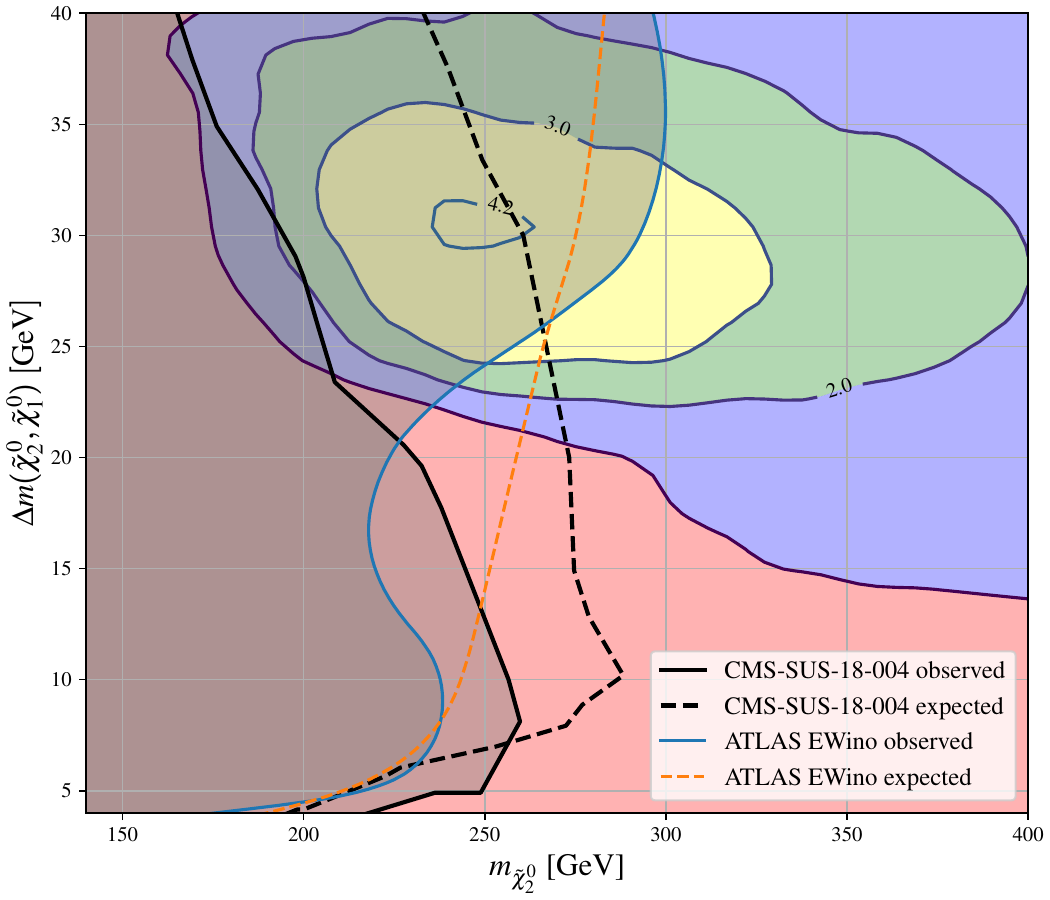}  
  \caption{\label{FIG:CMSonly}Results for only the CMS soft-lepton search in the wino-bino scenario, overlaid with soft-lepton excluded regions reported by ATLAS and CMS. Left: Likelihood ratio contours for the ``full'' simplified likelihood. Right: Likelihood ratio contours for the alternative simplified likelihood.}
\end{figure}

\begin{figure}
\captionsetup{width=0.85\textwidth}
\centering
  \includegraphics[width=0.485\textwidth]{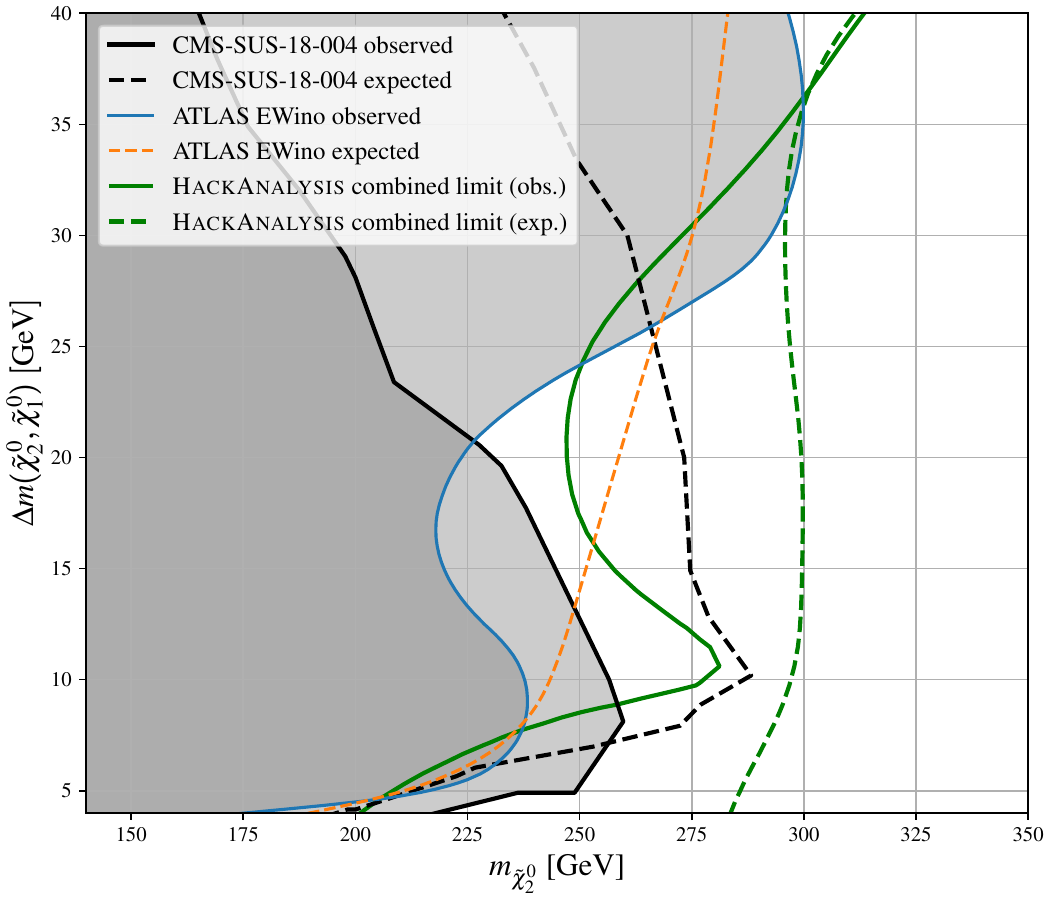}\\[0.5cm]
  \includegraphics[width=0.485\textwidth]{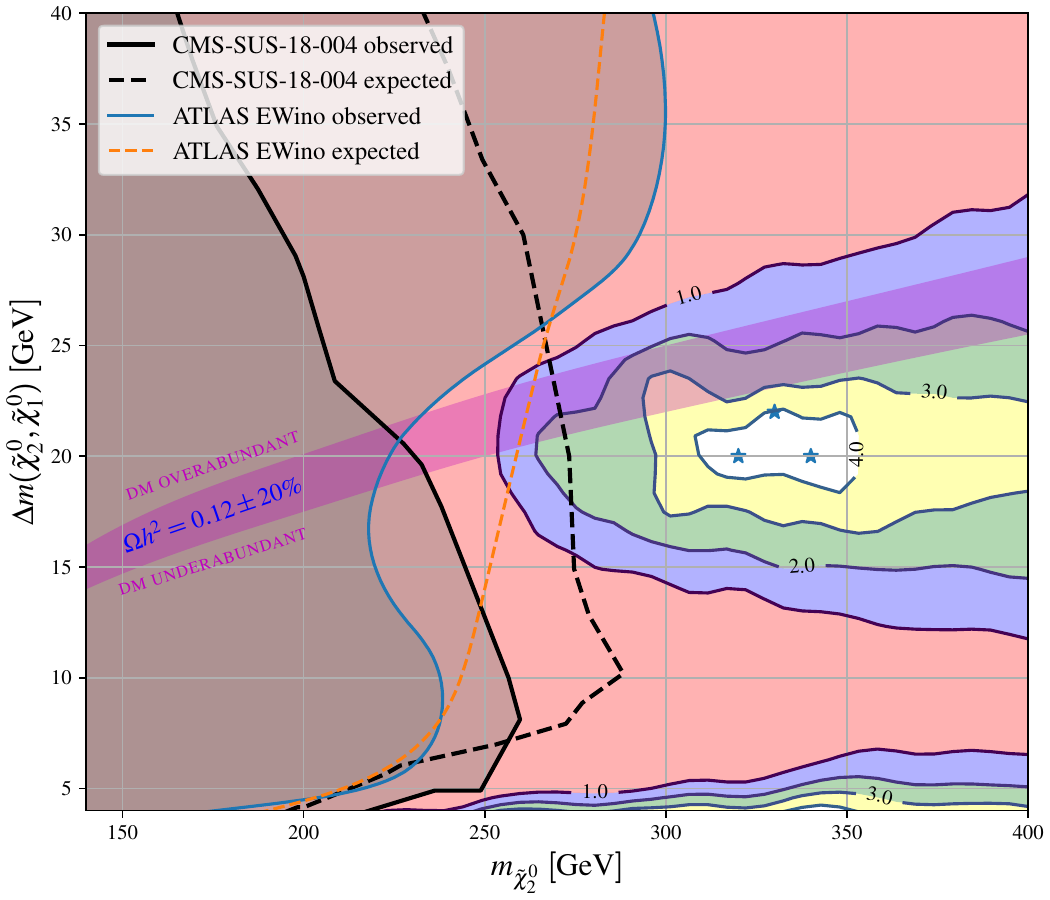} \hfill
  \includegraphics[width=0.485\textwidth]{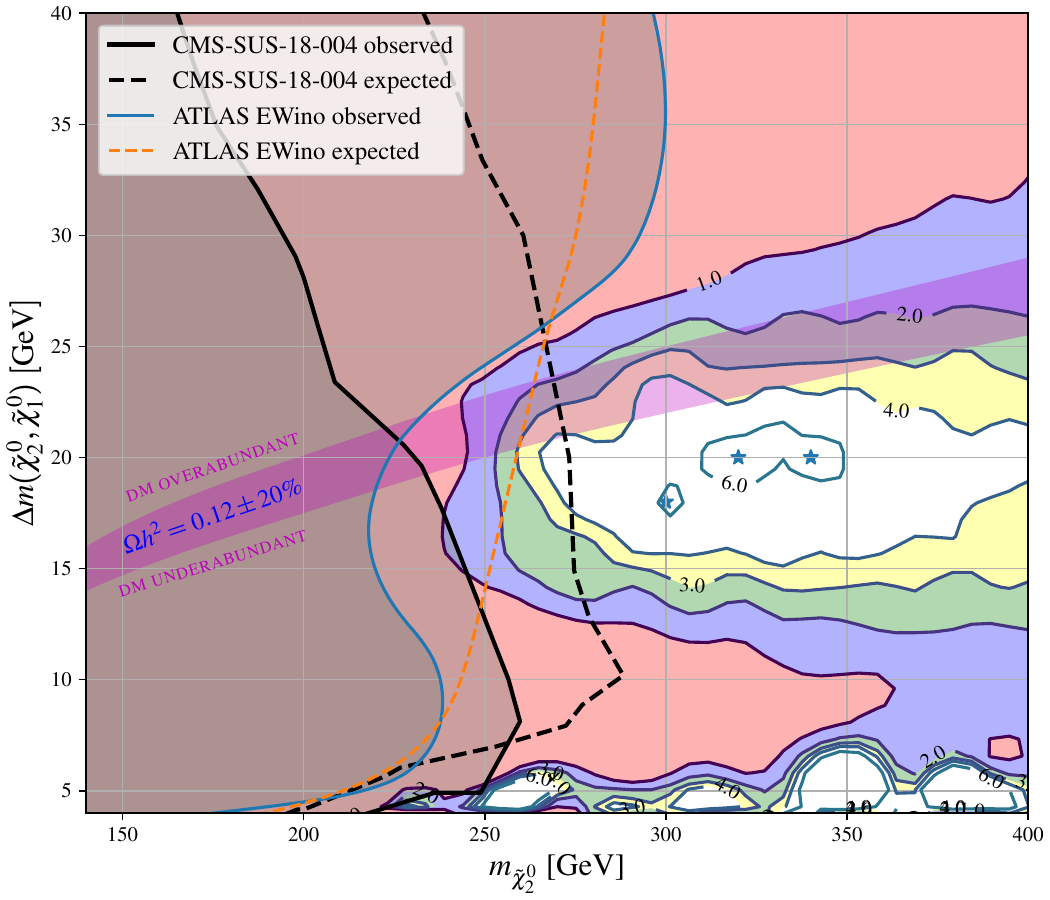} 
  \caption{\label{FIG:ALL}Results for all considered searches in the wino-bino scenario, overlaid with soft-lepton excluded regions reported by ATLAS and CMS. Top: Expected and observed limits using the ``full'' simplified likelihood. Bottom: Contours of Bayes factor $B_{10}$ using the full likelihood (left) and the alternative simplified one (right), highlighting the region of wino-bino parameter space capable of producing bino DM with a relic abundance $\Omega h^2 = 0.12 \pm 20\%$ through freeze out.}
\end{figure}

In Figure~\ref{FIG:CMSonly} we show the Bayes factor contours for (only) the CMS soft-lepton search for both the full (left panel) and alternative (right panel) simplified likelihoods (see Section~\ref{sec:CMSValidation}). The optimal region for that excess is in a region excluded by ATLAS, but there is a very wide region of parameter space where the excess is still manifest and not excluded; this overlaps with the ATLAS and monojet regions too. In the alternative simplified likelihood it can be seen that the $B_{10}$ values are larger and the best-fit region extends further beyond the region excluded by ATLAS. 
Hence we can examine the combination of all of our analyses (since all of them are orthogonal). We do this for both the full simplified likelihood and the alternative one. In the upper plot of Figure~\ref{FIG:ALL} we present the expected and observed combined limits, which are of course stronger than those from any individual search. The combination excludes additional parameter space lying between the expected and observed limits of ATLAS/CMS which naively might have seemed suitable for the excess; but leaves instead a wide region centred around $m_{\tilde{\chi}_2^0} \sim 275$~GeV and $\Delta m \sim 20$~GeV lying between the new observed and expected limits. From the lower plots, we observe that this is in fact the tip of the preferred region, the area of parameter space favoured with respect to the Standard Model by all searches. Altogether, we clearly see for the first time from this plot that the data favours a region around $m_{\tilde{\chi}_2^0} \sim 330$ GeV and $\Delta m \sim 20$ GeV.

At the same time, we overlay a band in this parameter space in which a (nearly) pure bino $\tilde{\chi}^0_1$ acts as dark matter with the correct relic abundance $\Omega h^2 \approx 0.12$. We discuss the calculation that produced this band in detail in Section~\ref{SEC:DM}. The fact that this passes very near to or through the best-fit region in both plots is a remarkable result: for the first time we are able to show that the combined data from compressed electroweakino and monojet searches may be pointing us to the values of the mass and properties of a dark matter particle. In anticipation of our discussion in Section~\ref{SEC:DM}, we note that the corresponding $\Omega h^2 \approx 0.12$ contour shown in the ATLAS analysis paper \cite{ATLAS:2021moa} does not have quite the same location as ours, indeed passing even more squarely through our best-fit region; the true location of the band may be shifted by higher-order corrections that we are unable to take into account, and also by the presence of the other supersymmetric particles that we have assumed to be integrated out. But we put off further discussion until Section~\ref{SEC:DM}.


\begin{table}\centering 
\captionsetup{width=0.85\textwidth}
  \begin{tabular}{c c c c c c c}
  \toprule
  \toprule
     & \multicolumn{4}{c}{$B_{10}$ (full)} & & \\ 
     \cline{2-5}
     \rule{0pt}{3ex}$(m_{\tilde{\chi}_2^0},\Delta m)$~[GeV] & ATLAS & CMS & CMS & Combined & $p$-value & $\hat{\mu}$\\
      & soft leptons & soft leptons & monojet & &  \\
    \midrule
    $330,22$ & $1.57$ & $1.40$ & $2.0$ & $4.46$ & $0.04$ & $1.09$\\
    $340,20$ & $2.24$ & $0.96$ & $2.3$ & $4.90$ & $0.03$ & $1.20$ \\
    $320,20$ & $2.54$ & $0.85$ & $2.3$ & $4.90$ & $0.04$ & $0.96$ \\
    \midrule
    $240,32$ & $0.003$  & $3.70$  & $1.35$ & $0.01$ & $0.40$ & $0.09$ \\
    $270,20$ & $2.80$ & $0.40$ & $2.0$ & $2.46$ & $0.04$ & $0.60$ \\                              
    $220,4$\,\,\, & $2.98$ & $0.085$ & $5.1$ & $1.30$ & $0.06$ & $0.50$ \\
    \bottomrule\bottomrule
  \end{tabular}\vspace{.85cm}

  \begin{tabular}{c c c c c c c}
  \toprule
  \toprule
     & \multicolumn{4}{c}{$B_{10}$ (alternative)} & & \\ 
     \cline{2-5}
     \rule{0pt}{3ex}$(m_{\tilde{\chi}_2^0},\Delta m)$~[GeV] & ATLAS & CMS & CMS & Combined & $p$-value & $\hat{\mu}$\\
      & soft leptons & soft leptons & monojet & &  \\
    \midrule
   $300,18$ & $2.50$ & $1.02 $ & $2.6 $ & $6.6 $ & $ 0.026 $ & $0.94$ \\
    $330,22$ & $1.57$ & $1.85$ & $2.0$ & $5.9$ & $0.026$ & $1.30$\\
   $340,20$ & $2.24$ & $1.30$  & $2.3 $ & $6.7 $ & $0.020$ & $1.46 $ \\
    $320,20$ & $2.54$ & $1.26$ & $2.3$ & $7.3 $ & $ 0.020$ & $1.17 $ \\
    \midrule
    $240,32$ & $0.003$  & $4.37$  & $1.35$ & $0.016 $ & $0.360 $   & $0.10$ \\
 $270,20$ & $2.80$ & $0.80$ & $2.00$ & $4.5$ & $0.028$ & $0.65$ \\            
    $220,4$\,\,\, & $2.98$ & $0.06$ & $5.1$ & $ 0.9 $ & $0.080$ & $0.45$ \\
    \bottomrule
    \bottomrule
  \end{tabular}

\caption{\label{TAB:bayes} Likelihood ratios $B_{10}$ for the best fit and selected other points, using the ``full'' simplified likelihood (top table) and the alternative one (lower table) for the CMS soft-lepton analysis. We give the factors for the two ATLAS soft-lepton analyses combined (``ATLAS soft leptons''), for the CMS soft-lepton analysis recast here and for the CMS monojet search, along with a combined factor (which is the product of the three), a local $p$-value for the point, and the best-fit signal strength $\hat{\mu}$.}
\end{table}

\subsection{Best-fit points}

In the upper panel of Table~\ref{TAB:bayes} we show the masses and likelihood ratios $B_{10}$ for the three best-fit points that can be seen in the lower left plot in Figure~\ref{FIG:ALL}, along with three additional interesting points; in the lower panel of the table we show the equivalent points for the alternative simplified likelihood seen in the lower right plot of Figure~\ref{FIG:ALL}. By showing the breakdown of the different Bayes factors as well as the combined value (which is of course the product of the other three), we can inspect the agreement of the four different analyses (we have combined the two ATLAS soft-lepton analyses into one column). With the full simplified likelihood, the point $m_{\tilde{\chi}_{2}^0} = 330$~GeV, $\Delta m = 22$~GeV is favoured compared to the SM by all of the analyses; while the points at $m_{\tilde{\chi}_{2}^0} = \{320, 240\}$~GeV, $\Delta m = 20$~GeV have slightly higher total Bayes factors despite being slightly disfavoured by the CMS soft-lepton analyses. But this last statement is too pessimistic: with the alternative simplified likelihood, which fits the exclusion data from CMS better, we find a best-fit point of $m_{\tilde{\chi}_{2}^0} = 320$ GeV, $\Delta m = 20$ GeV with an impressive $B_{10} = 7.3$ and values of $B_{10} > 1$ for all analyses considered.

We also list in the two tables some different interesting points that are individually compatible with one or two searches, but overall show small or no preference compared to the SM. The point at $(m_{\tilde{\chi}_{2}^0}, \Delta m) = (240,32)$~GeV is shown because it is the best-fit point for the CMS soft-lepton search with the full simplified likelihood, with $B_{10} = 3.7$. This point is however so strongly disfavoured by the ATLAS soft-lepton searches as to be excluded overall. The point at $ (m_{\tilde{\chi}_{2}^0}, \Delta m) = (270,20)$~GeV is the best-fit point for the ATLAS soft-lepton searches, but is somewhat disfavoured by CMS soft leptons for both full and alternative likelihoods. The final point $(m_{\tilde{\chi}_{2}^0}, \Delta m) = (220,4)$~GeV is a curious one, because it appears that the ATLAS soft-lepton searches display an excess at very small $\Delta m$, in the region favoured by monojets. If we just took the monojet searches and the ATLAS soft leptons, we would have a combined $B_{10} \simeq 15$. However, this appears strongly disfavoured by the CMS searches, although not to the point of excluding it, so there may well be points favoured at low $\Delta m$. It would be very interesting if the experimental collaborations could confirm or refute this finding.

\subsection{Dark matter and the realism of the simplified wino-bino(+) scenario}
\label{SEC:DM}

As mentioned above, the lower plots in Figure~\ref{FIG:ALL} each show a magenta band traversing the $(m_{\tilde{\chi}^0_2},\Delta m)$ parameter space. This band corresponds to a region in which a (nearly; see below) pure bino LSP is cosmologically stable and freezes out, during a standard cosmological history until it achieves a present-day abundance of $\Omega h^2 \approx 0.12$ consistent with the value favoured by the Planck collaboration following measurements of the power spectrum of the cosmic microwave background \cite{Planck:2018vyg}. Our band has some width, corresponding to allowing the DM relic abundance to vary by 20\% in either direction from the central Planck value. The bino LSP in this parameter space tends to achieve the correct relic abundance via coannihilations with the wino(s), which drives the fairly small favoured neutralino splitting and explains why points below the band feature underabundant (too-efficiently annihilating) dark matter.

We are motivated to include this band in the figures because it passes quite close by the points we highlighted in those figures, which, no matter which likelihood we consider, seem to fit all considered analyses quite well compared to the Standard Model. The interest to the community of a supersymmetric model with fairly light cold dark matter \emph{that could explain overlapping excesses at the LHC} is immediately obvious. But our DM results need to be explained and interpreted carefully, and they beg a discussion regarding how (and how seriously) we should take the simplified scenarios used by ATLAS and CMS.

We first note that the order of operations described in Section \ref{sec:CMSValidation} for event generation in the wino-bino(+) scenario to validate the CMS-SUS-18-004 recast does not straightforwardly allow us to calculate the DM relic abundance. This is because those event samples are generated on a rectangular grid in the $(m_{\tilde{\chi}^0_2},\Delta m)$ plane following manual modification of the physical electroweakino masses from parameter cards uploaded to \textsc{HepData} by the ATLAS collaboration for ATLAS-SUSY-2019-09 (the combined 2/3$\ell$ ATLAS soft-lepton analysis). This modification is done without performing a true calculation of the MSSM mass spectrum and, crucially, without computing the neutralino and chargino mixing matrices; to be clear, these mixings are left untouched, so the wino-bino(+) model is, in some sense, a fake model. This state of affairs is acceptable for the LHC (re)analysis because the electroweakino decay patterns are fixed without regard for any ``true'' mixing matrix: the production and decay always proceeds via off-shell $Z/W$ bosons. But it is inadequate for any discussion of dark matter, because the DM (co)annihilation rates depend sensitively on the composition of the electroweakinos, as do the limits from direct searches for dark matter which are quite sensitive to light higgsinos.

To explore this issue, there are two approaches. The first is to modify the parameter cards by hand so as to provide inputs to dark matter codes. In doing so, we can compute a dark matter relic abundance and direct detection constraints and find that the dark matter abundance is largely independent of the mixing, depending almost entirely on just $(m_{\tilde{\chi}_2^0}, \Delta m)$ as we would hope while the direct detection constraints can be varied over many orders of magnitude, essentially down to almost zero, by varying the mixing matrices.

By contrast, the second method involves mapping ``realistic'' MSSM model spaces onto the $(m_{\tilde{\chi}^0_2},\Delta m)$ parameter plane characteristic of the simplified models. To accurately explore the dark matter phenomenology in this manner, we turned again to \BSMArt, but this time we performed a number of Markov chain Monte Carlo (MCMC) scans over the fundamental parameters of the full MSSM. We used the MSSM implementation in \textsc{SARAH}~\cite{Staub:2008uz,Staub:2013tta,Goodsell:2014bna,Goodsell:2017pdq} to generate Fortran code, using routines from the \textsc{SPheno} library~\cite{Porod:2003um,Porod:2011nf}, to compute fermion masses including one-loop corrections and leading-order decays, including three-body decays. Moreover, we generated leading-order model files compatible with \textsc{CalcHEP}~\cite{Belyaev:2012qa} and hence \textsc{micrOMEGAs}~\cite{Belanger:2010pz, Belanger:2013oya, Alguero:2023zol}. Our scans finally called the latter program to compute the DM relic abundance and a $p$-value for DM direct detection (DD), with $p \leq 0.05$ indicating DD exclusion.

   \renewcommand{\arraystretch}{1.1}
\begin{table}
\captionsetup{width=0.85\textwidth}
\centering
    \begin{tabular}{l@{\hspace{3ex}}c@{\hspace{3ex}}}
    \toprule
    \toprule
Parameter & Value\\
    \midrule
$M_A^2$~[GeV$^2$] & $2.50 \times 10^7$\\
$\tan \beta$ & $25$\\
$M_3$~[GeV] & $588.3$\\
$\text{diag}(M_{Q,u,d}^2)$~[GeV$^2$] & $(8.25, 8.25, 0.125) \times 10^8$\\
$\text{diag}(M_{L,e}^2)$~[GeV$^2$] & $(4.25, 4.25, 0.0425) \times 10^8$\\
$A_{\tau}$~[GeV] & $-25.4$\\
$A_b$~[GeV] & $-110.7$\\
    \bottomrule
    \bottomrule
    \end{tabular}
    \caption{\label{tab:RealWinoBinoParams} Fixed inputs for MCMC scans over pseudo-simplified wino-bino MSSM parameter space.}
\end{table}
\renewcommand{\arraystretch}{1.0}

Our first scans began in parameter space resembling the wino-bino(+) SLHA cards uploaded to \textsc{HepData} by ATLAS. We immediately found that the light electroweakinos produced from ATLAS' values ($\mu$, $A_t$, etc.) contained quite a lot of higgsino and were often ruled out by direct detection, and anyhow the mass spectrum did not include a 125-GeV Higgs boson $h$ once two-loop corrections were included. We were therefore unable to quickly reproduce the $\Omega h^2 = 0.12$ contour shown in Figure 16(b) of ATLAS-SUSY-2019-09 for the wino-bino(+) scenario \cite{ATLAS:2021moa}. We then performed our own model simplification by fixing most MSSM parameters according to Table~\ref{tab:RealWinoBinoParams}, and scanned over the bino and wino masses $M_1$ and $M_2$. Our scans take the light gaugino masses from the ranges
\begin{align}
    M_1 \in [100,500]~\text{GeV}\ \ \ \text{and}\ \ \ M_2 - M_1 \in [-200,1000]~\text{GeV}.
\end{align}
Meanwhile, we experimented with various values of $\mu$ and $A_t$, requiring $m_h = (125.0 \pm 0.5)$~GeV, to better understand \emph{how decoupled} the higgsino must become in order to create a ``true'' wino-bino(+) scenario safe from direct detection. We ultimately find that $(\mu,A_t) = (600, 1300)$~GeV introduces the largest higgsino admixture that can be tolerated without ruling out the entire $(m_{\tilde{\chi}^0_2},\Delta m)$ plane on the basis of DM direct detection. In this scan output, we found points with the correct DM relic abundance being marginally ruled out by direct detection; these points feature LSP higgsino admixtures as high as 2.2\%. Values of $\mu$ higher than 600~GeV appear to uniformly produce points with LSP higgsino admixtures low enough to escape direct detection while showing the correct relic abundance, though some underabundant points continue to be excluded by DD until $\mu \gtrsim 1$~TeV.

In the course of this study, we also found (unsurprisingly) that the location of the $\Omega h^2 \approx 0.12$ band along the $\Delta m$ axis is mildly sensitive to the higgsino decoupling. In particular, heavier $\mu$ corresponds to smaller required $\Delta m$, presumably because the efficient light higgsino annihilations become kinematically suppressed and one has to rely on bino-wino coannihilations. Quantitatively, the band falls by 6--7~GeV along the $\Delta m$ axis as $\mu$ is raised from 500~GeV -- close to the value in ATLAS' wino-bino(+) SLHA cards for which DD rules out all points -- to 2~TeV ($A_t = 2$~TeV as well), by which point the higgsino component is reduced so much that we are (a) safe from direct detection and (b) firmly in the simplified-model regime. In this parameter space, it is straightforward to choose $M_1,M_2$ such that the LSP is well over 99.8\% bino and $\tilde{\chi}^{\pm}_1,\tilde{\chi}^0_2$ are nearly pure wino and split primarily by radiative corrections. This is the scenario in which we produce the bands in the two lower plots in Figure~\ref{FIG:ALL}, because it should be as close as practicable to the simplified wino-bino(+) scenario superimposed in those figures. 

We feel obligated to note, however, that our $\Omega h^2$ band does not agree exactly with the band shown in the ATLAS analysis \cite{ATLAS:2021moa}. The method used to compute that band is not described by ATLAS, instead quoting only \cite{Duan:2018rls}, which uses very similar tools to the ones we use, and with which we agree (see Figure 5 of that reference). It would be interesting to have clarification on this point. Perhaps some private computation was provided, potentially including higher-order effects not available in \micromegas. In the end, this exercise confirms that the simplified supersymmetric wino-bino(+) model used by the experimental collaborations to interpret the soft-lepton excesses can be reconstructed from the full MSSM, but such reconstruction is actually \emph{required} in order to see a global (DM-inclusive) picture of the model.

\section{Conclusions}
\label{sec:Conclusion}

We have presented the details of recasting two new searches relevant for electroweakinos and used them to examine the compatibility of the wino-bino simplified model scenario. In the process, we demonstrated the compatibility of the excesses, showing the overlap of the regions of parameter space where the data are fit better by the model than the Standard Model. By completing the set of available Run 2 analyses most relevant for low-energy electroweak-charged particles with compressed spectra, we were also able to place new constraints on the CFDM model. Our most striking result is that the preferred region passes near, if not through, the band for the observed dark matter relic abundance, potentially predicting the mass and properties of a dark matter particle! This allows us to point to where new particles are most likely to be found in Run 3.

Our work showed the limitations of simplified likelihoods in describing models with excesses: there are some aspects of our new recast of the CMS soft-lepton search for which we could not understand the discrepancy with the reported experimental limits, but this was alleviated by combining some regions. It would be very useful if we could have input from the collaboration, and ideally a {\sc COMBINE} statistical model.

Otherwise, this work now paves the way for studies of different models. We saw in \cite{Agin:2024yfs} that it can be challenging to find models that explain the excesses well, but that paper was far from exhaustive. Indeed, since the ATLAS search SUSY-2018-16 found a $p$-value of $0.0034$ for an unconstrained higgsino template, whereas the wino-bino model for the same search yields only a modest Bayes factor, we imagine another model may yield even better combined significances than we found here.  We also eagerly anticipate the release of a full statistical model for the ATLAS monojet search EXOT-2018-06, so that the significance of those searches can be included in similar future studies. Finally, a \madanalysis implementation of the CMS soft-lepton search will be available soon, so that it can be used to provide limits along with the library of other searches in the Public Analysis Database.

\section*{Acknowledgments}

B. F. and M. D. G. are supported in part by Grant ANR-21-CE31-0013, Project DMwithLLPatLHC, from the \emph{Agence Nationale de la Recherche} (ANR), France. We thank Jack Y. Araz for collaboration on closely related topics and interesting discussions, especially about the use of \spey to combine likelihoods; and Krzysztof Rolbiecki for interesting discussions.

\bibliographystyle{jhep}
\bibliography{literature}

\end{document}